\documentclass[a4paper,11pt]{article}
\pdfoutput=1 

\usepackage{jheppub} 

\usepackage{graphicx,color,amsmath,bm}

\usepackage[T1]{fontenc} 

\newcommand{\be}{\begin{equation}}
\newcommand{\ee}{\end{equation}}
\newcommand{\bea}{\begin{eqnarray}}
\newcommand{\eea}{\end{eqnarray}}

\newcommand{\bfk}{\mbox{\boldmath $k$}}

\def\kt{k_\perp}
\def\pt{p_\perp}
\newcommand{\bfp}{\mbox{\boldmath $p$}}

\newcommand{\bfq}{\mbox{\boldmath $q$}}

\newcommand{\bfS}{\mbox{\boldmath $S$}}

\newcommand{\bfL}{\mbox{\boldmath $L$}}

\def\pp{p_\perp}
\newcommand{\pup}{p^\uparrow}

\newcommand{\pdown}{p^\downarrow}

\def\avk{\langle k_\perp ^2\rangle}
\def\avp{\langle p_\perp ^2\rangle}
\def\avkS{\langle k_S ^2\rangle}

\def\lsim{\mathrel{\rlap{\lower4pt\hbox{\hskip1pt$\sim$}}\raise1pt\hbox{$<$}}}
\def\gsim{\mathrel{\rlap{\lower4pt\hbox{\hskip1pt$\sim$}}\raise1pt\hbox{$>$}}}
\def\nostrocostruttino#1\over#2{\mathrel{\mathop{\kern 0pt \rlap
{\hbox{$#1$}}} \hbox{\kern-.135em $#2$}}}

\preprint{JLAB-THY-16-2404}

\title{\boldmath Study of the sign change of the Sivers function from STAR Collaboration W/Z production data}

\author[a,b]{M. Anselmino,}
\author[a,b]{M. Boglione,}
\author[c,d]{U. D'Alesio}
\author[d]{F. Murgia}
\author[e,f]{A. Prokudin}

\affiliation[a]{Dipartimento di Fisica Teorica, Universit\`a di Torino,
                Via P.~Giuria 1, I-10125 Torino, Italy}
\affiliation[b]{INFN, Sezione di Torino, Via P.~Giuria 1, I-10125 Torino, Italy}
\affiliation[c]{Dipartimento di Fisica, Universit\`a di Cagliari, Cittadella
Universitaria, I-09042 Monserrato (CA), Italy}
\affiliation[d]{INFN, Sezione di Cagliari, Cittadella Universitaria, I-09042
Monserrato (CA), Italy}
\affiliation[e]{Science Division, Penn State University Berks, Reading, Pennsylvania 19610, USA}
\affiliation[f]{Theory Center, Jefferson Lab, 12000 Jefferson Avenue, Newport News, VA 23606, USA}

\emailAdd{mauro.anselmino@to.infn.it}
\emailAdd{elena.boglione@to.infn.it}
\emailAdd{umberto.dalesio@ca.infn.it}
\emailAdd{francesco.murgia@ca.infn.it}
\emailAdd{prokudin@jlab.gov}
\abstract{
Recent data on the transverse single spin asymmetry $A_N$ measured by the STAR
Collaboration for $\pup \, p \to W^\pm/Z^0 \, X$ reactions at RHIC allow the
first investigation of the Sivers function in Drell-Yan processes and of its
expected sign change with respect to SIDIS processes. A new extraction of the
Sivers functions from the latest SIDIS data is performed and a critical
assessment of the significance of the STAR data is attempted.
}

\begin{document}

\maketitle

\flushbottom

\section{\label{Intro} Introduction}

The Transverse Momentum Dependent Partonic Distribution Functions (TMD-PDFs)
encode  information on the 3-dimensional structure of nucleons in momentum
space; they depend on the parton intrinsic motion inside the nucleon and, in
general, on the nucleon and parton spins. At leading twist there are eight
independent TMD-PDFs which have been studied in Semi Inclusive Deep Inelastic
Scattering (SIDIS) processes. Among them, the Sivers distribution, which
describes the momentum distribution of unpolarised quarks and gluons inside
a transversely polarised proton, has a clear experimental
signature~\cite{Airapetian:2009ae,Adolph:2012sp} and is of particular interest for several reasons; one expects
it to be related to fundamental intrinsic features of the nucleon and to basic
QCD properties.

In fact, the Sivers distribution $\Delta^N \! f_ {q/\pup}$ relates the
motion of unpolarised quarks
and gluons to the nucleon spin $\bfS$; then, in order to build a scalar,
parity invariant quantity, $\bfS$ must couple to the only other available
pseudo-vector, that is the parton orbital angular momentum, $\bfL_q$ or
$\bfL_g$. Another peculiar feature of the Sivers distribution is that its
origin at partonic level can be traced in QCD interactions between the quarks
(or gluons) active in inelastic high energy interactions and the nucleon
remnants~\cite{Brodsky:2002cx,Brodsky:2002rv}; thus, it is expected to be
process dependent and have opposite signs in SIDIS and Drell-Yan (D-Y)
processes~\cite{Collins:2002kn,Brodsky:2013oya}:
\be
\Delta^N \! f_ {q/\pup}(x,\kt)|_{\rm SIDIS} = - \Delta^N \!
f_ {q/\pup}(x,\kt)|_{\rm D-Y}. \label{eq:sign-change}
\ee
This important prediction remains to be tested.

The Sivers distribution can be accessed through the study of azimuthal
asymmetries in polarised SIDIS and Drell-Yan (D-Y) processes. These have
been clearly observed in the last years, in SIDIS, by the
HERMES~\cite{Airapetian:2009ae}, COMPASS~\cite{Adolph:2012sp} and Jefferson
Lab \cite{Allada:2013nsw} Collaborations, allowing  extractions
of the SIDIS Sivers function~\cite{Anselmino:2008sga,Bacchetta:2011gx,
Anselmino:2012aa,Echevarria:2014xaa}. However, no information could be obtained
on the D-Y Sivers function, as no polarised D-Y process had ever been measured.

Recently, first  data from polarised D-Y processes at RHIC, $\pup \,p \to
W^\pm/Z^0 \, X$, have become available~\cite{Adamczyk:2015gyk}. The data
show an azimuthal asymmetry, $A_N^W$, which can be interpreted as due to the
Sivers effect and which hints ~\cite{Adamczyk:2015gyk,Huang:2015vpy} at a sign
change between the Sivers function observed in these D-Y processes and the
Sivers function extracted from SIDIS processes. However, considering the
importance of the sign change issue, before drawing any definite conclusion,
both the SIDIS and D-Y data and their comparison, have to be critically
analysed and discussed.

In this paper we perform a new extraction of the valence and sea-quark Sivers
functions from the newest experimental SIDIS data. We then perform an analysis
of the RHIC $W^\pm/Z^0$ D-Y data~\cite{Adamczyk:2015gyk}, based on these
new functions, trying to assess the significance of $A_N^W$ on
the sign change of the Sivers functions.

The paper is organised as follows: in Section~\ref{Form} we recall the
formalism used to analyse and interpret the experimental data. In
Section~\ref{Extr} we present a new extraction of the Sivers functions
from experimental data. In Section~\ref{Pred} we compute the asymmetries
observable in D-Y processes and based on the SIDIS extracted Sivers functions,
both with and without the sign change, comparing them with the recent RHIC
results, while in Section~\ref{Imp} we analyse the impact of the D-Y data as
a possible indication of the sign change of the Sivers function. Conclusions
and final comments are given in Section~\ref{Com}.

\section{\label{Form} Formalism}

We consider a generalised Drell-Yan process, $\pup\,p \to W^\pm\,X$, in
which one observes a $W$ boson, with four-momentum $q$, created by the
annihilation of a quark and an antiquark. We define our kinematical
configuration with the polarised $\pup$ proton, with four-momentum
$p_1$, moving along the positive $z$-axis and the unpolarised one, with
four-momentum $p_2$, moving opposite to it. We adopt the usual variables:
\be
q = (q_0, \bfq_T, q_L) \quad\quad q^2 = M_{_W}^2 \quad\quad
y_{_W} = \frac 12 \ln \frac{q_0 + q_L}{q_0 - q_L} \quad \quad
x_F = \frac{2\,q_L}{\sqrt s}\quad\quad s=(p_1 + p_2)^2 \>.
\label{var}
\ee

The annihilating quarks have an intrinsic transverse motion, $\bfk_{\perp 1}$
and $\bfk_{\perp 2}$. We fix the azimuthal angles by choosing the ``up"
$(\uparrow)$ polarisation direction as the positive $y$-axis ($\phi_S = \pi/2$).
The spin ``down"$(\downarrow)$ polarisation direction will have $\phi_S = 3\pi/2$.
The other transverse momenta azimuthal angles are defined as:
\be
\bfq_T = q_T(\cos\phi_{_W}, \, \sin\phi_{_W},\, 0) \quad\quad\quad
\bfk_{\perp i} = k_{\perp i}(\cos\varphi_i, \, \sin\varphi_i, \, 0)
\quad\quad (i = 1,2) \>. \label{phi-dep}
\ee

In the kinematical region
\be
q_T^2 \ll M_{_W}^2 \quad\quad\quad k_{\perp} \simeq q_T \>, \label{kinr}
\ee
using the TMD factorisation formalism at leading order, the unpolarised
cross section for the $p\,p \to W\,X$ process can be written
as~\cite{Anselmino:2009st,Kang:2009bp,Huang:2015vpy,Peng:2014hta}
\be
\frac{d \sigma^{pp \to WX}}{dy_{_W} \, d^2\bfq_T} =
\hat{\sigma}_0 \sum_{q_1, q_2} |V_{q_1, q_2}|^2
\int d^2\bfk_{\perp 1} \, d^2\bfk_{\perp 2} \>
\delta^2(\bfk_{\perp 1} + \bfk_{\perp 2} - \bfq_T) \>
f_{q_1/p}(x_1, k_{\perp 1}) \>
f_{q_2/p}(x_{2}, k_{\perp 2}) \>,
\ee
where $f_{q_i/p}(x_i, k_{\perp i})$ are the unpolarised TMDs,
$V_{q_1, q_2}$ are the weak interaction CKM matrix elements and the
$\sum_{q_1, q_2}$ runs over all appropriate light quark and antiquark flavours
($q_1q_2 = u\bar d, \bar du, u\bar s, \bar su$ for $W^+$, etc.).
$\hat{\sigma}_0$ is the lowest-order partonic cross section (with $G_F$
the Fermi weak coupling constant),
\be
\hat{\sigma}_0 = \frac{\sqrt 2 \, \pi \, G_F \, M_{_W}^2}{3\,s} \>,
\ee
and the parton longitudinal momentum fractions are given, at
${\cal O}\left({\kt}/{M_{_W}} \right)$, by
\be
x_{1,2} = \frac{M_{_W}}{\sqrt s} \, e^{\pm y_{_W}}  =
\frac{\pm x_F + \sqrt{x_F^2 + 4 \, M_{_W}^2/s}}{2} \> \cdot \label{xqxqb}
\ee
Notice that, with the definition of $x_F$ adopted in Eq.~(\ref{var}),
one has
\be
x_F = x_1 - x_2 \quad\quad\quad\quad\quad |x_F| \leq 1 -
\frac{M_{_W}^2}{s} \> \cdot
\ee

In such a formalism, the distribution for unpolarised quarks with
transverse momentum $\bfk_\perp$ inside a proton with 3-momentum $\bfp$
and spin $\bfS$,
\bea
\hat f_ {q/\pup} (x,\bfk_\perp) &=& f_ {q/p} (x,\kt) +
\frac{1}{2} \, \Delta^N \! f_ {q/\pup}(x,\kt)  \;
{\bfS} \cdot (\hat {\bfp}  \times
\hat{\bfk}_\perp) \nonumber \\
&=& f_ {q/p} (x,\kt) - \frac{k_\perp}{m_p} \>
f_{1T}^{\perp q}(x, k_\perp) \;
{\bfS} \cdot (\hat {\bfp}  \times \hat{\bfk}_{\perp}) \>,
\eea
generates a transverse Single Spin Asymmetry (SSA)
\bea
&& \hspace*{-1.2cm} A_N^W = \frac{d\sigma^{\pup p \to W X}
          - d\sigma^{\pdown p \to W X}}
            {d\sigma^{\pup p \to W X}
          + d\sigma^{\pdown p \to W X}}
\equiv \frac{d\sigma^\uparrow - d\sigma^\downarrow}
           {d\sigma^\uparrow + d\sigma^\downarrow} \> ,\label{asy} \\
&& \hspace*{-1.2cm} d\sigma^\uparrow - d\sigma^\downarrow = \hat{\sigma}_0\sum_{q_1, q_2} |V_{q_1, q_2}|^2
\int \! d^2\bfk_{\perp 1} \, d^2\bfk_{\perp 2} \>
\delta^2(\bfk_{\perp 1} + \bfk_{\perp 2} - \bfq_T) \> \nonumber \\
&& \hspace*{3.3cm} \times \> {\bfS} \cdot (\hat {\bfp}_1 \times \hat{\bfk}_{\perp 1}) \>
\Delta^N\!f_{q_1/\pup}(x_1,k_{\perp 1}) \>
f_{q_2/p}(x_2,k_{\perp 2}) \> ,\\
&& \hspace*{-1.2cm} d\sigma^\uparrow + d\sigma^\downarrow = 2 \hat{\sigma}_0 \sum_{q_1, q_2}  |V_{q_1, q_2}|^2
\int \! d^2\bfk_{\perp 1} \, d^2\bfk_{\perp 2} \>
\delta^2(\bfk_{\perp 1} + \bfk_{\perp 2} - \bfq_T) \> \nonumber \\
&& \hspace*{3.3cm}\times \>
f_{q_1/p}(x_1,k_{\perp 1}) \>
f_{q_2/p}(x_2,k_{\perp 2}) \> .
~ \label{ann}
\eea
where $d \sigma$ stands for $d \sigma^{pp \to WX}/(dy_{_W} \, d^2\bfq_T)$
and $\Delta^N\!f_{q/\pup}(x, k_{\perp})$ is the Sivers function.

The above expression much simplifies adopting, as usual, a Gaussian factorised
form both for the unpolarised distribution and the Sivers functions, as in
Ref.~\cite{Anselmino:2008sga}:
\bea
f_{q/p}(x,k_\perp) &=& f_q(x) \, \frac{1}{\pi \langle\kt^2\rangle} \,
e^{-{\kt^2}/{\langle\kt^2\rangle}} \ ,\label{partond}\\
\Delta^N \! f_{q/\pup}(x,\kt) &=& 2 \, {\cal N}_q(x) \,
h(\kt) \, f_ {q/p} (x,\kt) \ , \label{sivfac} \\
{\cal N}_q(x) &=& N_q\ x^{\alpha_q} (1-x)^{\beta_q} \ \frac{(\alpha_q + \beta_q)^{(\alpha_q + \beta_q)}}{\alpha_q^{\alpha_q} \beta_q^{\beta_q}} \ , \label{siversn} \\
h(\kt) &=& \sqrt{2e}\,\frac{k_\perp}{M_{1}}\,e^{-{k_\perp^2}/{M_{1}^2}}\> ,
\label{siverskt}
\eea
where $f_q(x)$ are the unpolarised PDFs, $M_1$ is a parameter which allows
the $k_\perp$ Gaussian dependence of the Sivers function to be different from
that of the unpolarised TMDs and ${\cal N}_q(x)$ is a function which
parameterises the factorised $x$ dependence of the Sivers function.

The following moment of the Sivers function is of importance:
\bea
\Delta^N \! f_{q/\pup}^{(1)}(x) = \int d^2 \bfk_\perp \frac{\kt}{4 m_p}
\Delta^N \! f_{q/\pup}(x,\kt) = - f_{1T}^{\perp (1) q}(x) \, , \label{siversm1} \\
\Delta^N \! f_{q/\pup}^{(1)}(x) = \frac{\sqrt{\frac{e}{2}} \ \langle \kt^2 \rangle M_1^3}{m_p (\langle \kt^2 \rangle + M_1^2)^2}  \ {\cal N}_q(x)  f_q(x)
\label{siversm2} \ .
\eea

With the choices of Eqs.~(\ref{partond})-(\ref{siverskt}) the $\bfk_\perp$
integrations can be performed analytically in Eq.~(\ref{ann}), obtaining:
\bea
A_N^W(y_{_W}, \bfq_T) &=& {\bfS} \cdot (\hat {\bfp}_1  \times {\hat\bfq}_{T})
\, \frac{2\,\avkS^2}{[\avkS + \avk]^2}
\> \exp\Biggl[ - \frac{q_T^2}{2\,\avk} \, \Biggl(\frac{\avk - \avkS}
{\avk + \avkS} \Biggr) \Biggr]
\frac{\sqrt{2\,e}\,q_T}{M_1} \nonumber \\
&\times& \frac
{\sum_{q_1, q_2} |V_{q_1, q_2}|^2 \, {\cal N}_{q_1}(x_1) f_{q_1}(x_1) \>
f_{q_2}(x_{2})}
{\sum_{q_1, q_2} |V_{q_1, q_2}|^2 \, f_{q_1}(x_1) \> f_{q_2}(x_{2})}\label{Anan} \\
&\equiv& \cos\phi_{_W} \, A_N(y_{_W}, q_T) \label{Ameas}
\eea
with
\be
\avkS = \frac{ M_1^2 \, \langle k_{\perp}^2\rangle}
{M_1^2 + \langle k_{\perp}^2\rangle}
\ee
and where, in the last line, we have used, according to our kinematics,
${\bfS} \cdot (\hat {\bfp}_1  \times \hat{\bfq}_{T}) = \cos\phi_{_W}$.
$A_N(y_{_W}, q_T)$ is the quantity measured at RHIC~\cite{Adamczyk:2015gyk}
\footnote{Notice that in Ref.~\cite{Adamczyk:2015gyk} there is a deceptive
definition of $\cos\phi$, which is opposite to ours. However, we have
checked with the STAR Collaboration that the quantity measured is exactly
that defined in Eq.~(\ref{Ameas}).}.

Let us notice that the RHIC measurements of $W^\pm$ production at $\sqrt{s}
= 500$ GeV~\cite{Adamczyk:2015gyk} cover the rapidity region $|y_{_W}|<1$.
In particular, data are available for $y_{_W} \simeq \pm \, 0.4$ and
$y_{_W} \simeq 0$.
This corresponds to:
\bea
y_{_W} \simeq -0.4 \quad & x_1 \simeq 0.11 \quad & x_2 \simeq 0.24 \nonumber \\
y_{_W} \simeq 0 \quad & x_1 \simeq 0.16 \quad & x_2 \simeq 0.16 \label{eq:kin}\\
y_{_W} \simeq +0.4 \quad & x_1 \simeq 0.24 \quad & x_2 \simeq 0.11  \>, \nonumber
\eea
where $x_1$ refers to the polarised proton and $x_2$ to the unpolarised one.
Then, although the $x$ region is predominantly the valence one, the data at
$y_{_W} \simeq -0.4$ are expected to be more sensitive to the
sea-quark Sivers functions.

\section{\label{Extr} Extraction of Sivers functions from SIDIS data}

The quark flavours involved in $W$ production include anti-quarks. Thus,
in order to estimate the asymmetry $A_N^W$, it is important to have a reliable
extraction of both quark and anti-quark Sivers functions.

For instance, in order to produce a $W^+$, $u$, $\bar d$ and $\bar s$ quarks
from the polarised proton combine with $\bar d$, $\bar s$, $u$ quarks from
the unpolarised proton, such that the asymmetry is proportional to
\be
  |V_{u,d}|^2 \, \left( \Delta^N\!f_{u/\pup} \otimes f_{\bar d/p}
+ \Delta^N\!f_{\bar d/\pup} \otimes f_{u/p} \right)
+ |V_{u,s}|^2 \, \left( \Delta^N\!f_{u/\pup} \otimes f_{\bar s/p}
+ \Delta^N\!f_{\bar s/\pup} \otimes f_{u/p} \right).
\label{eq:u}
\ee
Both quantities in the round brackets in the above equation contain a sea
and a valence quark distribution. However, because of the numerical values
\footnote{$|V_{u, d}| = 0.97417 \pm0.00021$,
$|V_{u, s}| = 0.2248 \pm 0.0006$, from Ref.~\cite{Agashe:2014kda}.}
of $|V_{u, d}|$ and $|V_{u, s}|$,
the last two terms in Eq.~(\ref{eq:u}) are much suppressed with respect
to the first two. Thus, we expect that $A_N^{W^+}$ mainly
depends on the $u$ quark and $\bar d$ sea quark Sivers functions.

Likewise, for $W^-$ production, the asymmetry is proportional to
\be
  |V_{u,d}|^2 \, \left( \Delta^N\!f_{\bar u/\pup} \otimes f_{d/p}
+ \Delta^N\!f_{d/\pup} \otimes f_{\bar u/p} \right)
+ |V_{u,s}|^2 \, \left( \Delta^N\!f_{\bar u/\pup} \otimes f_{s/p}
+ \Delta^N\!f_{s/\pup} \otimes f_{\bar u/p} \right),
\ee
and we expect that $W^-$ data are mainly sensitive to $d$ quark and
$\bar u$ sea quark Sivers function.

A previous extraction of the Sivers functions that included anti-quark
distributions was reported in Ref.~\cite{Anselmino:2008sga}. However,
new data have become available since then and we perform here a new complete
extraction of the Sivers functions. We refer to Ref.~\cite{Anselmino:2008sga}
for more details about the procedure.

One may notice that in our simple parameterisation of the Sivers functions
as given in Eqs.~(\ref{partond})-(\ref{siverskt}) the knowledge of the width
$\langle\kt^2\rangle$ of the unpolarised TMDs is important. Such a study
was performed in Refs.~\cite{Anselmino:2013lza,Signori:2013mda}. We adopt
here the parameters from Ref.~\cite{Anselmino:2013lza}, fixed by fitting
the HERMES multiplicities~\cite{Airapetian:2012ki}:
\be
\langle \kt^2 \rangle = 0.57 \pm 0.08\; \textrm{GeV}^2  \hspace{1cm}
\langle \pp^2 \rangle = 0.12 \pm 0.01\; \textrm{GeV}^2 \;,
\label{hermes-par}
\ee
where $\langle \pp^2 \rangle $ is the width of unpolarised Transverse Momentum
Dependent Fragmentation Functions (TMD-FFs):
\bea
D_{h/q}(z,p_\perp) &=& D_{h/q}(z) \, \frac{1}{\pi \langle\pt^2\rangle} \,
e^{-{\pt^2}/{\langle\pt^2\rangle}} \>. \label{partonff}
\eea
Notice that the study of Ref.~\cite{Anselmino:2013lza} found no flavour
dependence of the widths of the TMDs. The collinear distribution and
fragmentation functions, $f_{q/p}(x)$ and $D_{h/q}(z)$, needed for our
parameterisations are taken from the available fits of the world data: in
this analysis we use the CTEQ6L set for the PDFs~\cite{Pumplin:2002vw}
and the DSS set for the fragmentation functions~\cite{deFlorian:2007aj}.
The LHAPDF~\cite{Buckley:2014ana} library is used for collinear PDFs.
We fit the latest data from the HERMES Collaboration on the SIDIS Sivers
asymmetries for $\pi^\pm$ and $K^\pm$ production off a proton
target~\cite{Airapetian:2009ae}, the COMPASS Collaboration data on
LiD \cite{Alekseev:2008aa} and  NH$_3$ targets \cite{Adolph:2014zba},
and JLab data on $^3$He target~\cite{Qian:2011py}.

These available SIDIS data cover a relatively narrow region of $x$, typically
in the so-called valence region. It suffices to use the most simple
parameterisation for the anti-quark Sivers functions
[see Eqs.~(\ref{sivfac}), (\ref{siversn})]:
\bea
 {\cal N}_{\bar q}(x) &=& N_{\bar q} \label{siversnbar}  \ .
\eea
It means that we assume the anti-quark Sivers functions to be proportional to
the corresponding unpolarised PDFs; we have checked that a fit allowing for
more complicated
structures of Eq.~\eqref{siversn} for the anti-quarks, results in undefined
values of the parameters $\alpha$ and $\beta$.

The Sivers asymmetry measured in SIDIS can be expressed using our parameterisations
of TMD functions from Eqs.~(\ref{partond}-\ref{siverskt}, \ref{partonff}) as
\bea
A_{UT}^{\sin(\phi_h - \phi_S)}(x,y,z, P_T) &=&
\, \frac{ [z^2 \avk + \avp]\avkS^2}{[z^2 \avkS + \avp]^2 \avk}
\> \exp\Biggl[ - \frac{P_T^2 \, z^2 (\avkS - \avk)}{(z^2 \avkS + \avp)
(z^2 \avk + \avp)} \Biggr] \nonumber\\
&\times&
\frac{\sqrt{2\,e}\,z\,P_T}{M_1} 
\frac
{\sum_{q} e_q^2 \, {\cal N}_{q}(x) f_{q}(x) \>
D_{h/q}(z)}
{\sum_{q} e_q^2 \, f_{q}(x) \> D_{h/q}(z)} \; \cdot \label{ASIDIS}
\eea
%
%
\begin{figure}[t]
\centering
\includegraphics[width=16cm]{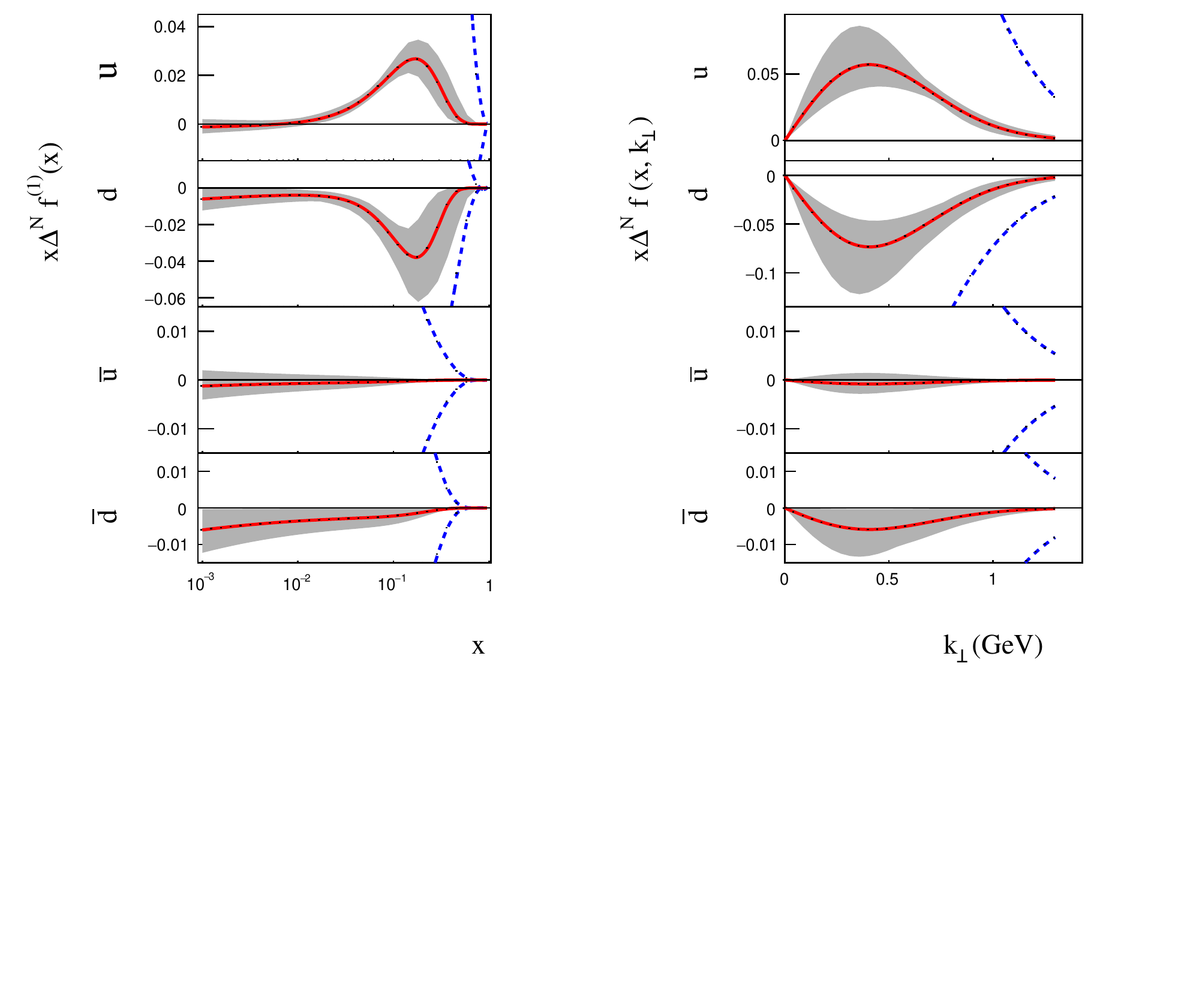}
\vskip -4cm
\caption{Extracted Sivers distributions for $u = u_v + \bar u$,
$d = d_v + \bar d$, $\bar u$ and $\bar d$ at $Q^2=2.4$ GeV$^2$.
Left panel: the first moment of the
Sivers functions, Eqs.~(\ref{siversm1}) and (\ref{siversm2}) of the text,
versus $x$. Right panel: plots of the Sivers functions, Eq.~(\ref{siversn})
of the text, at $x=0.1$ versus
$k_\perp$. The solid lines correspond to the best fit. The dashed lines
correspond to the positivity bound of the Sivers functions. The shaded
bands correspond to our estimate of 95\% C.L. error.}
\label{fig:functions}
\end{figure}
%
%
Thus, we introduce a total of 9 free parameters for valence and sea-quark
Sivers functions: $N_{u_v}$, $N_{d_v}$, $N_{\bar u}$, $N_{\bar d}$, $\alpha_u$,
$\beta_u$, $\alpha_d$, $\beta_d$, and $M_1^2$ (GeV$^2$). In order to estimate
the errors on the parameters and on the calculation of the asymmetries we
follow the Monte Carlo sampling method explained in Ref.~\cite{Anselmino:2008sga}.
That is, we generate samples of parameters $\bm{\alpha}_i$, where each
$\bm{\alpha}_i$ is
an array of random values of $\{N_{u_v},N_{d_v},N_{\bar u}, N_{\bar d}, \alpha_u,
\alpha_d,\beta_u,\beta_d,M_1^2\}$, in the vicinity of the minimum found by
MINUIT, $\bm{\alpha}_0$, that defines the minimal total $\chi^2$ value,
$\chi^2_{\rm min}$.
We generate $2\cdot 10^4$ sets of parameters $\bm{\alpha}_i$ that satisfy
\bea
\chi^2(\bm{\alpha}_i) \leq \chi^2_{\rm min} + \Delta \chi^2\, ,
\label{condition}
\eea
with the high tolerance $\Delta \chi^2 = 17.21$  that corresponds to the
95\% C.L. of coverage probability for 9 free parameters.
The fit is performed with MINUIT minimisation package and the resulting
parameters can be found in Table~\ref{parameters}; the corresponding extracted
Sivers functions are shown in Fig.~\ref{fig:functions}. We indicate both
the errors for the standard definition of $\Delta \chi^2 = 1$ and the high
tolerance error with $\Delta \chi^2 = 17.21$ (the errors given in parentheses).

The main new features of the fit are the parameters $N_{d_v} = - 0.52 \pm 0.20$
and $N_{u_v} = 0.18 \pm 0.04$. The previous extraction \cite{Anselmino:2008sga},
that used different gaussian width values, $\langle \kt^2 \rangle = 0.25\;
\textrm{GeV}^2$ and $\langle \pp^2 \rangle = 0.20 \;\textrm{GeV}^2$, yielded
$N_{d} = - 0.9$, which almost saturated the positivity bound $|N_q| = 1$, and
$N_{u} = 0.35$. The $\bar u$ and $\bar d$ Sivers functions turn out to be both
small, compared to the quark distributions, and negative. Future Electron-Ion
Collider data will be crucial for the investigation of the anti-quark Sivers
distributions. The parameters that control the large-$x$ behaviour of the
functions, $\beta_{u_v}$ and $\beta_{d_v}$, have big errors, see
Table~\ref{parameters}. The future Jefferson Lab 12 GeV data will allow a
better precision extraction in the high-$x$ region.

The partial contributions to $\chi^2$ from different experiments are shown in
Table~\ref{results_sidis}. One can see that the proton data on $\pi^+$ from the
HERMES Collaboration and the positive hadron data from the COMPASS Collaboration
show some larger $\chi^2$ values that might be attributed to possible effects
of TMD evolution~\cite{Aybat:2011ta,Anselmino:2012aa,Echevarria:2014xaa}.

\begin{figure}[t]
\centering
\includegraphics[width=7.0cm]{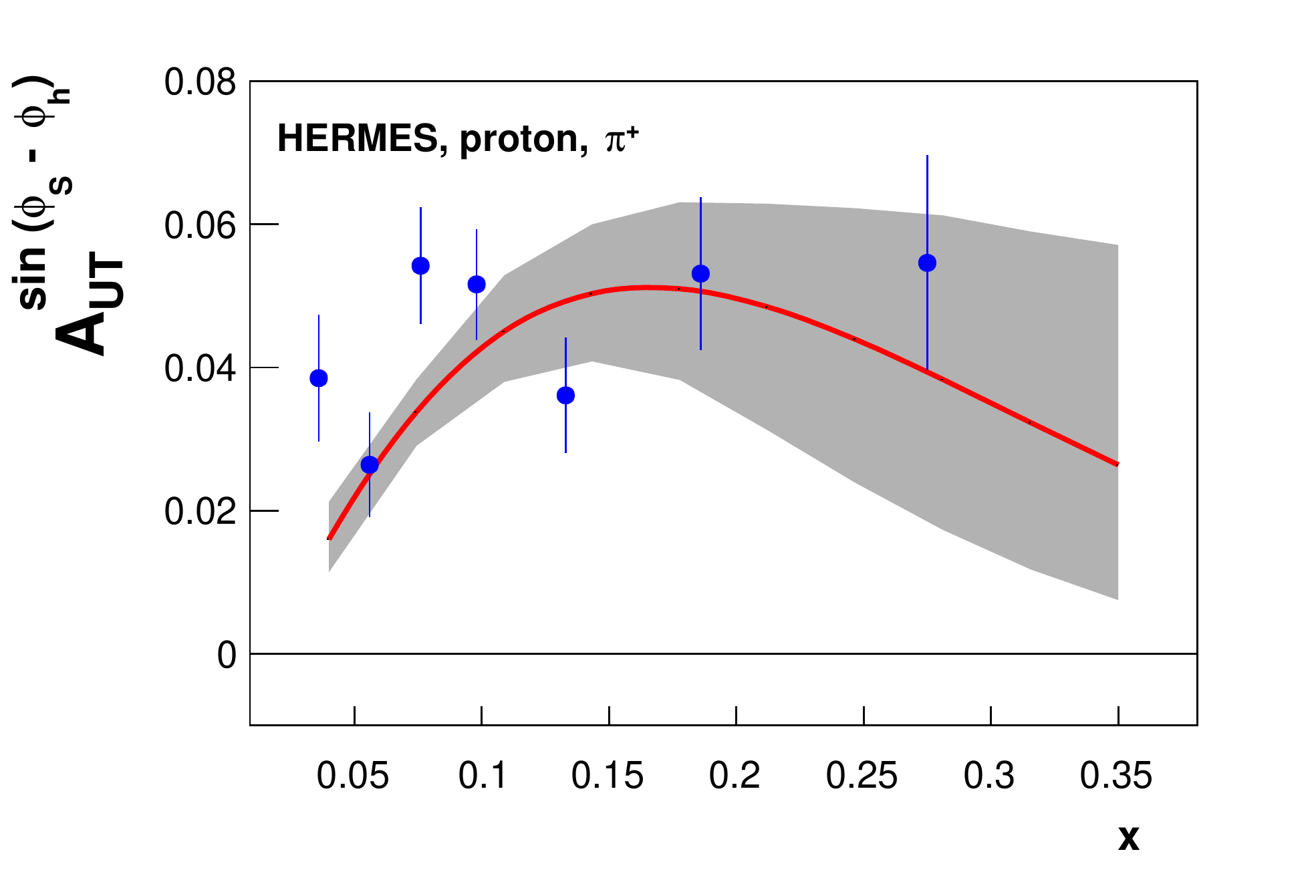}(a)
\includegraphics[width=7.0cm]{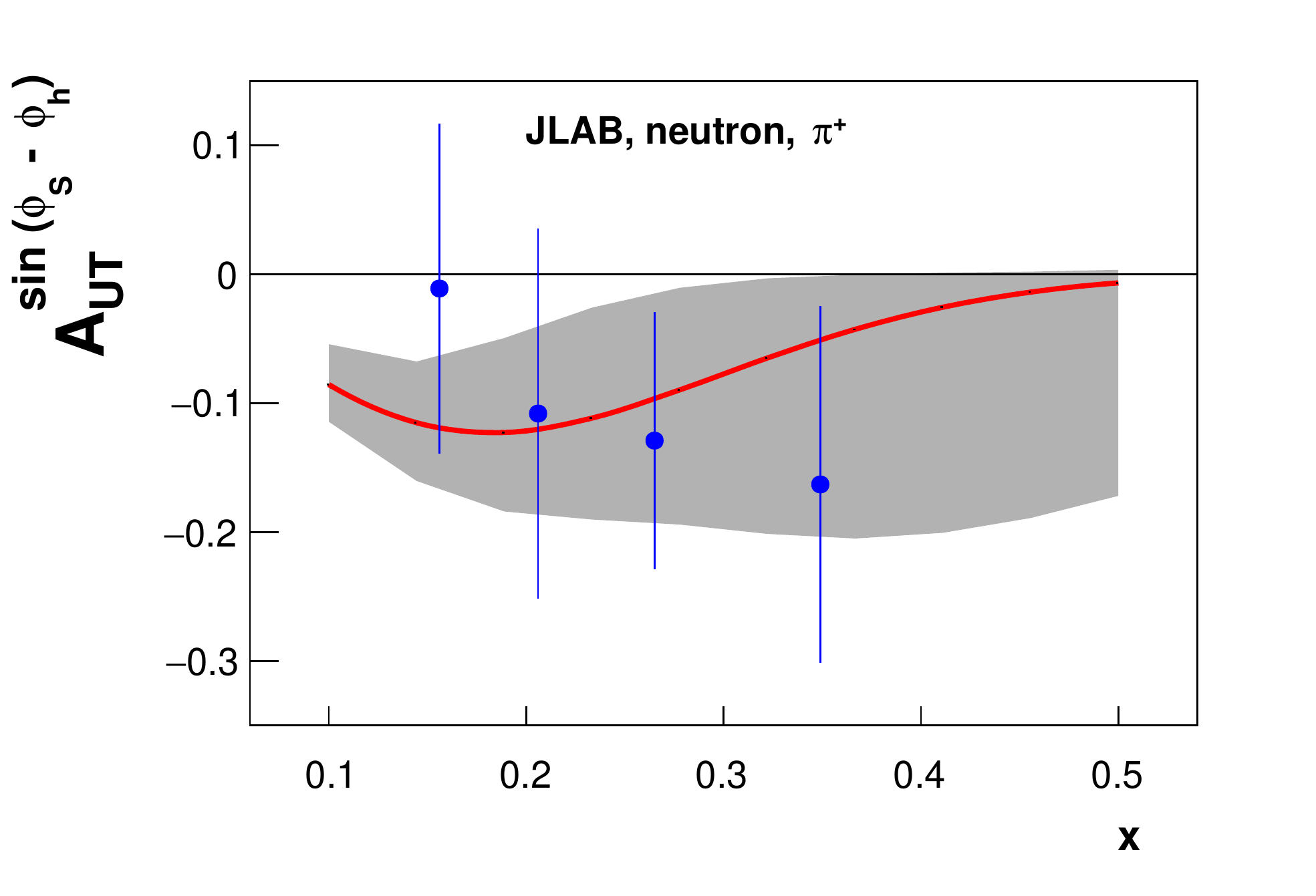}(b)
\includegraphics[width=7.0cm]{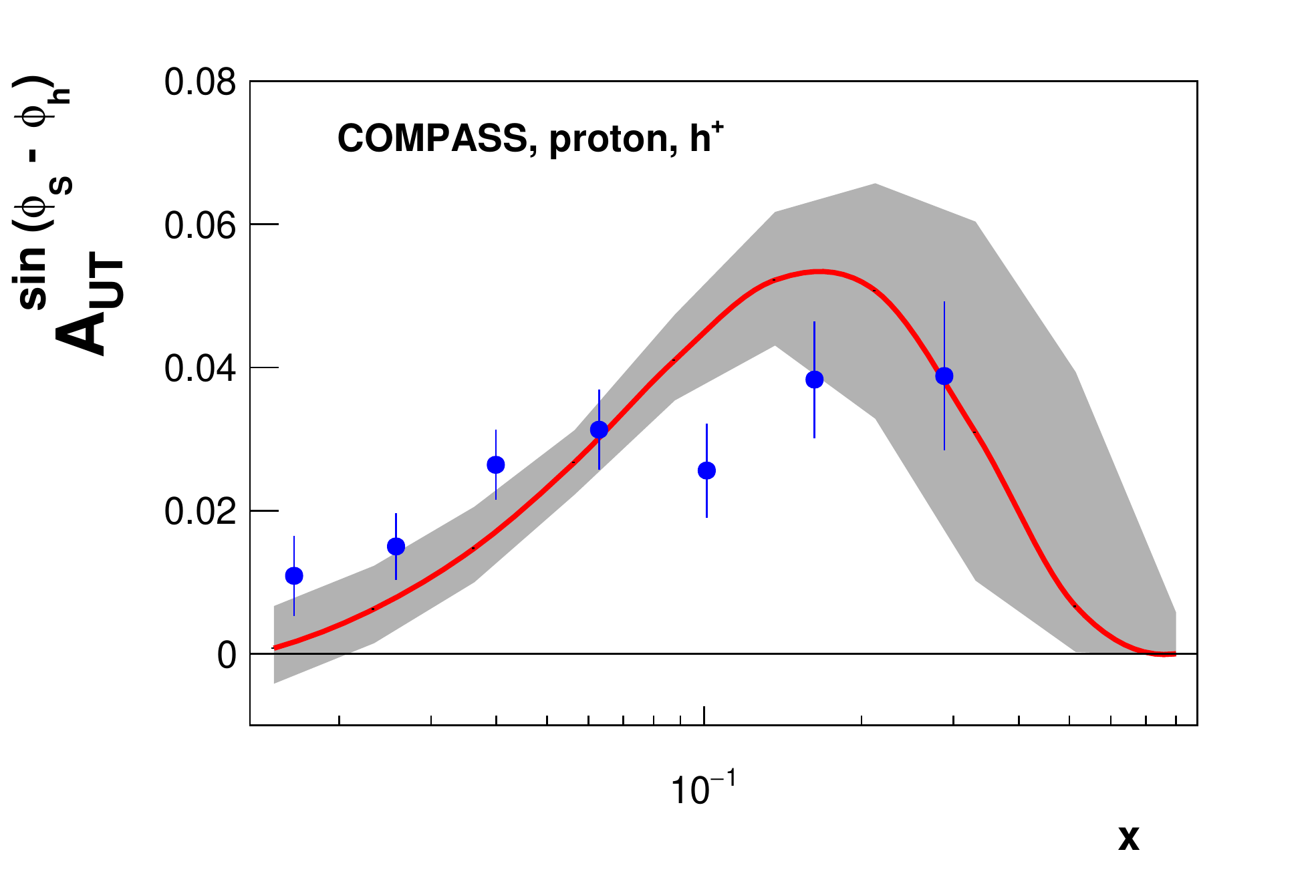}(c)
\includegraphics[width=7.0cm]{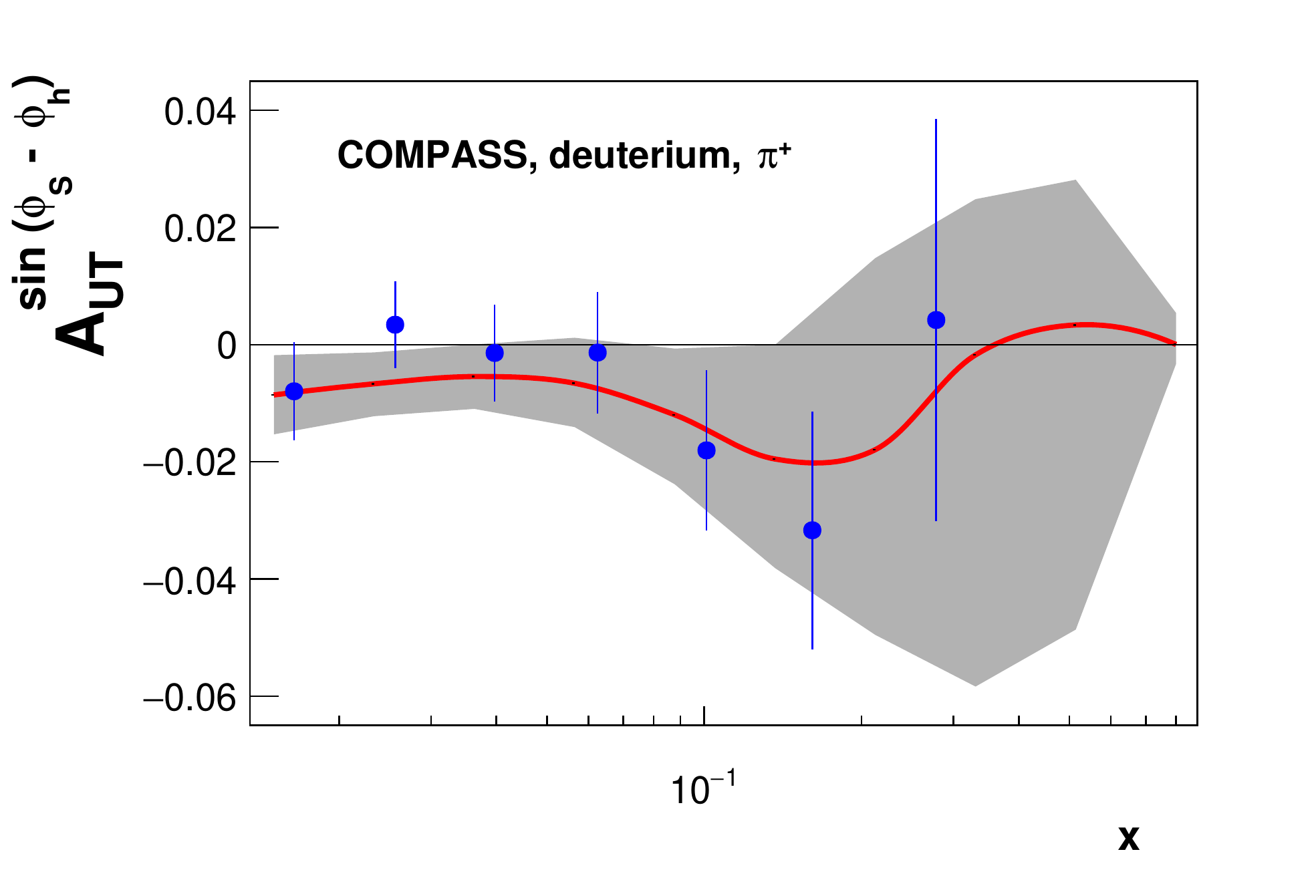}(d)
\caption{Examples of best fits of SIDIS experimental data:
(a) Data from the HERMES Collaboration for $\pi^+$ production off hydrogen
target as function of $x$.
(b) Data from JLab 6 for $\pi^+$ production off $^3$He target as function of $x$.
(c) Data from the COMPASS Collaboration for $h^+$ production off NH$_3$
target as function of $x$.
(d) Data from the COMPASS Collaboration for $\pi^+$ production off LiD
target as function of $x$.
The solid lines correspond to the best fit. The shaded region corresponds to our estimate of 95\% C.L. error band.}
\label{fig:hermes_jlab}
\end{figure}

Several plots showing the quality of our best fits of the data are presented in
Fig.~\ref{fig:hermes_jlab}.
\begin{table}[t]
\begin{tabular}{l c l  l c l  l c l}
\hline
\hline
\\
$N_{u_v}$ &=& $0.18\pm 0.01 (\pm 0.04)$ & $\alpha_{u_v}$ &=& $ 1.0 \pm 0.3 (\pm 0.6)$ & $\beta_{u_v}$ &=& $ 6.6 \pm 2.0 (\pm 5.2)$ \\
$N_{d_v}$ &=& $-0.52\pm 0.08 (\pm 0.20)$ & $\alpha_{d_v}$ &=& $ 1.9 \pm 0.5 (\pm 1.5)$ & $\beta_{d_v}$ &=& $ 10. \pm 4.0 (\pm 11.)$  \\
$N_{\bar u}$ &=& $-0.01\pm 0.01 (\pm 0.03)$  \\
$N_{\bar d}$ &=& $-0.06\pm 0.02 (\pm 0.06)$ \rule[-1em]{0pt}{1em}\\
\hline
\\
$M_1^2$ &=& $0.8\pm 0.2 (\pm 0.9)$ (GeV$^2$)\\
\multicolumn{3}{l}{$\chi_{\rm min}^2 =  325.29$} & \multicolumn{3}{l}
{$\chi^2_{\rm min}/{\rm dof}=1.29$}
\rule[-1.3em]{0pt}{1.0em}\\
\hline
\hline
\end{tabular}
\caption{Fitted parameters of the Sivers valence quark and anti-quark
distributions for $u_v$, $d_v$, $\bar u$, $\bar d$. The fit is performed by
using MINUIT minimisation package. Quoted errors correspond to MINUIT estimate
with $\Delta \chi^2 = 1$, and $\Delta \chi^2 = 17.21$ for errors in
parentheses.}
\label{parameters}
\end{table}
\newpage
\begin{table}[htb]
\begin{tabular}{l c c c c c c}
\hline
\hline
Experiment & Hadron & Target &  Dependence & ndata & $\chi^2$ &
$\chi^2/{\rm ndata}$ \\
\hline
\hline
JLAB \cite{Qian:2011py}  & $ \pi^+ $ & $^3$He & $x$ & 4 &     2.24 &     0.56  \\
JLAB \cite{Qian:2011py}  & $ \pi^- $ & $^3$He & $x$ & 4 &     3.50 &     0.87  \\
\hline
HERMES \cite{Airapetian:2009ae}  & $ \pi^0 $ & H & $x$ & 7 &     5.63 &     0.80  \\
HERMES \cite{Airapetian:2009ae}  & $ \pi^+ $ & H & $x$ & 7 &    18.72 &     2.67  \\
HERMES \cite{Airapetian:2009ae}  & $ \pi^- $ & H & $x$ & 7 &    14.82 &     2.12  \\
HERMES \cite{Airapetian:2009ae}  & $ \pi^0 $ & H & $z$ & 7 &     7.43 &     1.06  \\
HERMES \cite{Airapetian:2009ae}  & $ \pi^+ $ & H & $z$ & 7 &     4.26 &     0.61  \\
HERMES \cite{Airapetian:2009ae}  & $ \pi^- $ & H & $z$ & 7 &     4.60 &     0.66  \\
HERMES \cite{Airapetian:2009ae}  & $ \pi^0 $ & H & $P_T$ & 7 &     5.85 &     0.84  \\
HERMES \cite{Airapetian:2009ae}  & $ \pi^+ $ & H & $P_T$ & 7 &    17.13 &     2.45  \\
HERMES \cite{Airapetian:2009ae}  & $ \pi^- $ & H & $P_T$ & 7 &     6.62 &     0.95  \\
HERMES \cite{Airapetian:2009ae}  & $ K^+ $ & H & $x$ & 7 &     8.90 &     1.27  \\
HERMES \cite{Airapetian:2009ae}  & $ K^- $ & H & $x$ & 7 &     4.46 &     0.64  \\
HERMES \cite{Airapetian:2009ae}  & $ K^+ $ & H & $z$ & 7 &     9.94 &     1.42  \\
HERMES \cite{Airapetian:2009ae}  & $ K^- $ & H & $z$ & 7 &     8.49 &     1.21  \\
HERMES \cite{Airapetian:2009ae}  & $ K^+ $ & H & $P_T$ & 7 &     8.38 &     1.20  \\
HERMES \cite{Airapetian:2009ae}  & $ K^- $ & H & $P_T$ & 7 &     5.70 &     0.81  \\
\hline
COMPASS \cite{Alekseev:2008aa}  & $ \pi^+ $ & LiD & $x$ & 9 &     3.09 &     0.34  \\
COMPASS \cite{Alekseev:2008aa}  & $ \pi^- $ & LiD & $x$ & 9 &     4.75 &     0.53  \\
COMPASS \cite{Alekseev:2008aa}  & $ \pi^+ $ & LiD & $z$ & 8 &     6.30 &     0.79  \\
COMPASS \cite{Alekseev:2008aa}  & $ \pi^- $ & LiD & $z$ & 8 &    10.86 &     1.36  \\
COMPASS \cite{Alekseev:2008aa}  & $ \pi^+ $ & LiD & $P_T$ & 9 &     5.94 &     0.66  \\
COMPASS \cite{Alekseev:2008aa}  & $ \pi^- $ & LiD & $P_T$ & 9 &     4.65 &     0.52  \\
COMPASS \cite{Alekseev:2008aa}  & $ K^+ $ & LiD & $x$ & 9 &     8.13 &     0.90  \\
COMPASS \cite{Alekseev:2008aa}  & $ K^- $ & LiD & $x$ & 9 &    12.02 &     1.34  \\
COMPASS \cite{Alekseev:2008aa}  & $ K^+ $ & LiD & $z$ & 8 &     9.70 &     1.21  \\
COMPASS \cite{Alekseev:2008aa}  & $ K^- $ & LiD & $z$ & 8 &     9.39 &     1.17  \\
COMPASS \cite{Alekseev:2008aa}  & $ K^+ $ & LiD & $P_T$ & 9 &     6.40 &     0.71  \\
COMPASS \cite{Alekseev:2008aa}  & $ K^- $ & LiD & $P_T$ & 9 &    15.10 &     1.68  \\
\hline
COMPASS \cite{Adolph:2014zba}  & $ h^+ $ & NH$_3$ & $x$ & 9 &    33.76 &     3.75  \\
COMPASS \cite{Adolph:2014zba}  & $ h^- $ & NH$_3$ & $x$ & 9 &    12.14 &     1.35  \\
COMPASS \cite{Adolph:2014zba}  & $ h^+ $ & NH$_3$ & $z$ & 8 &    16.56 &     2.07  \\
COMPASS \cite{Adolph:2014zba}  & $ h^- $ & NH$_3$ & $z$ & 8 &    14.87 &     1.86  \\
COMPASS \cite{Adolph:2014zba}  & $ h^+ $ & NH$_3$ & $P_T$ & 9 &     8.29 &     0.92  \\
COMPASS \cite{Adolph:2014zba}  & $ h^- $ & NH$_3$ & $P_T$ & 9 &    12.41 &     1.38  \\
\hline
\hline
\end{tabular}
\caption{Partial $\chi^2$ values of the global best fit for SIDIS experiments.}
\label{results_sidis}
\end{table}
%

\section{\label{Pred} Predictions for W and Z asymmetries and comparison with data}

We can now compute the asymmetry $A_N(y_{_W}, q_T)$, according to
Eqs.~(\ref{Anan})-(\ref{Ameas}), using the Sivers functions -- {\it or
their opposite} -- as given in Eqs.~(\ref{partond})-(\ref{siverskt})
with the parameters, and the corresponding uncertainties, shown in
Table~\ref{parameters}.

\begin{figure}[htbp]
\centering
\includegraphics[width=7.cm]{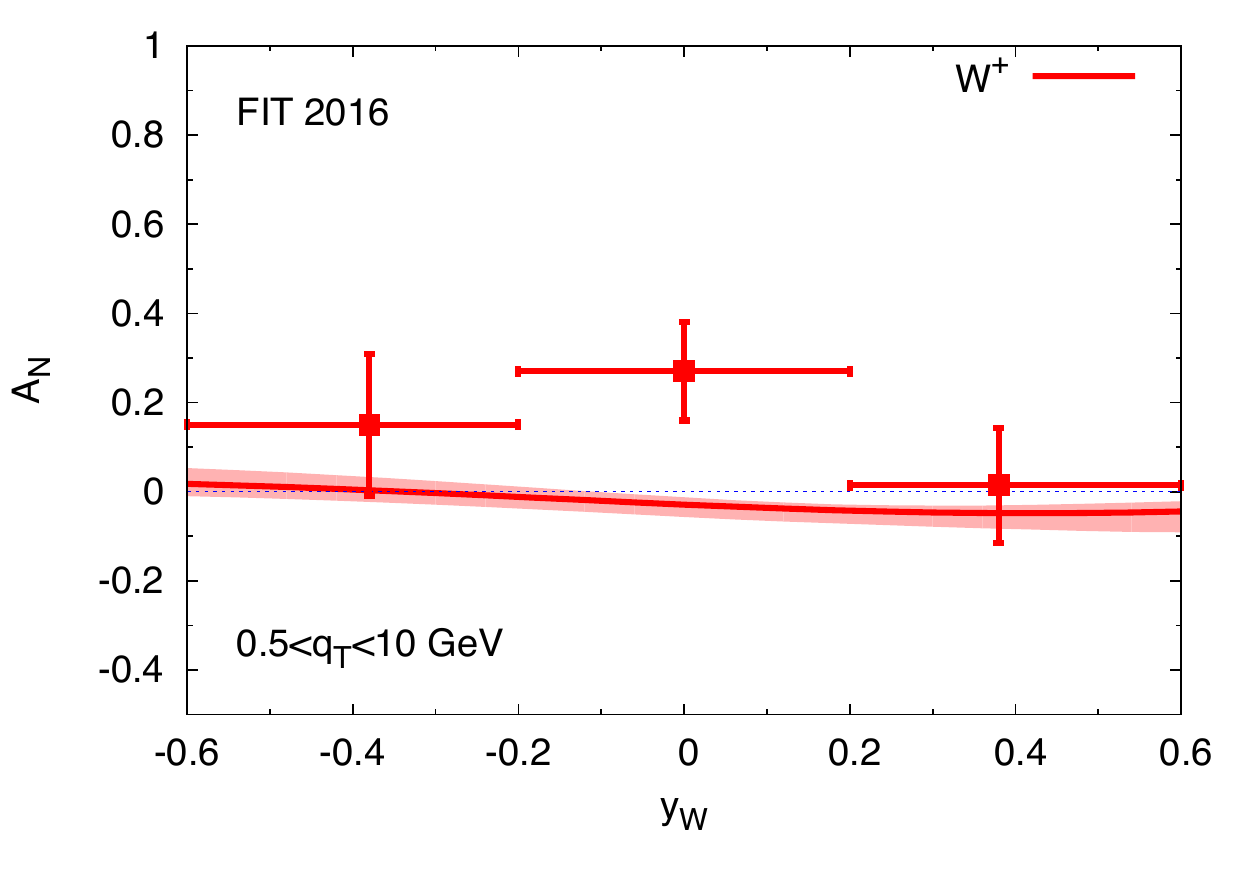}(a)
\includegraphics[width=7.cm]{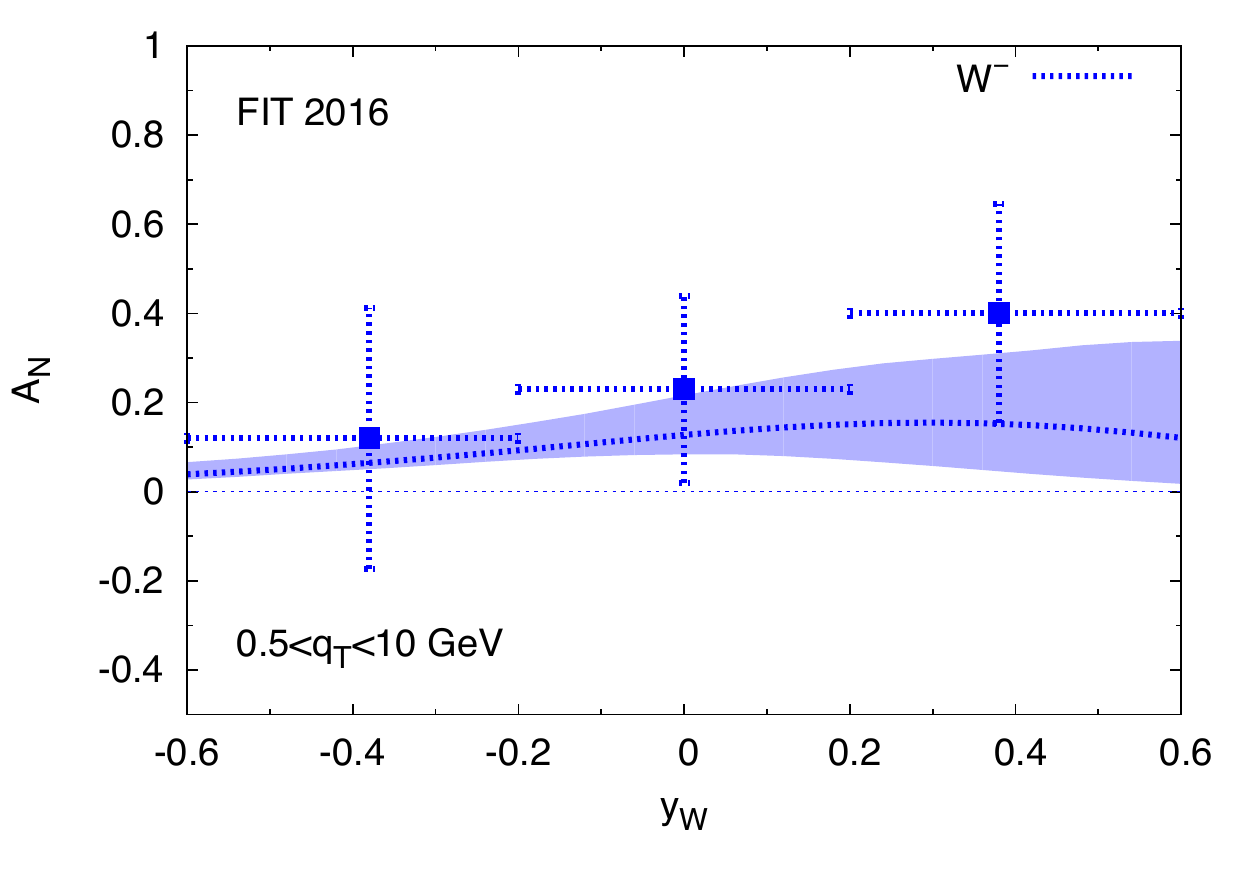}(b)
\caption{Our estimates of the Sivers asymmetry $A_N$ for $W^+$ (a) and $W^-$ (b)
production, assuming a sign change of the SIDIS Sivers functions, compared
with the experimental data as function of $y_{_W}$. $q_T$ is integrated in the
region $[0.5, 10]$ GeV.}
\label{fig:sivers_W}
\end{figure}

\begin{figure}[htbp]
\centering
\includegraphics[width=9cm]{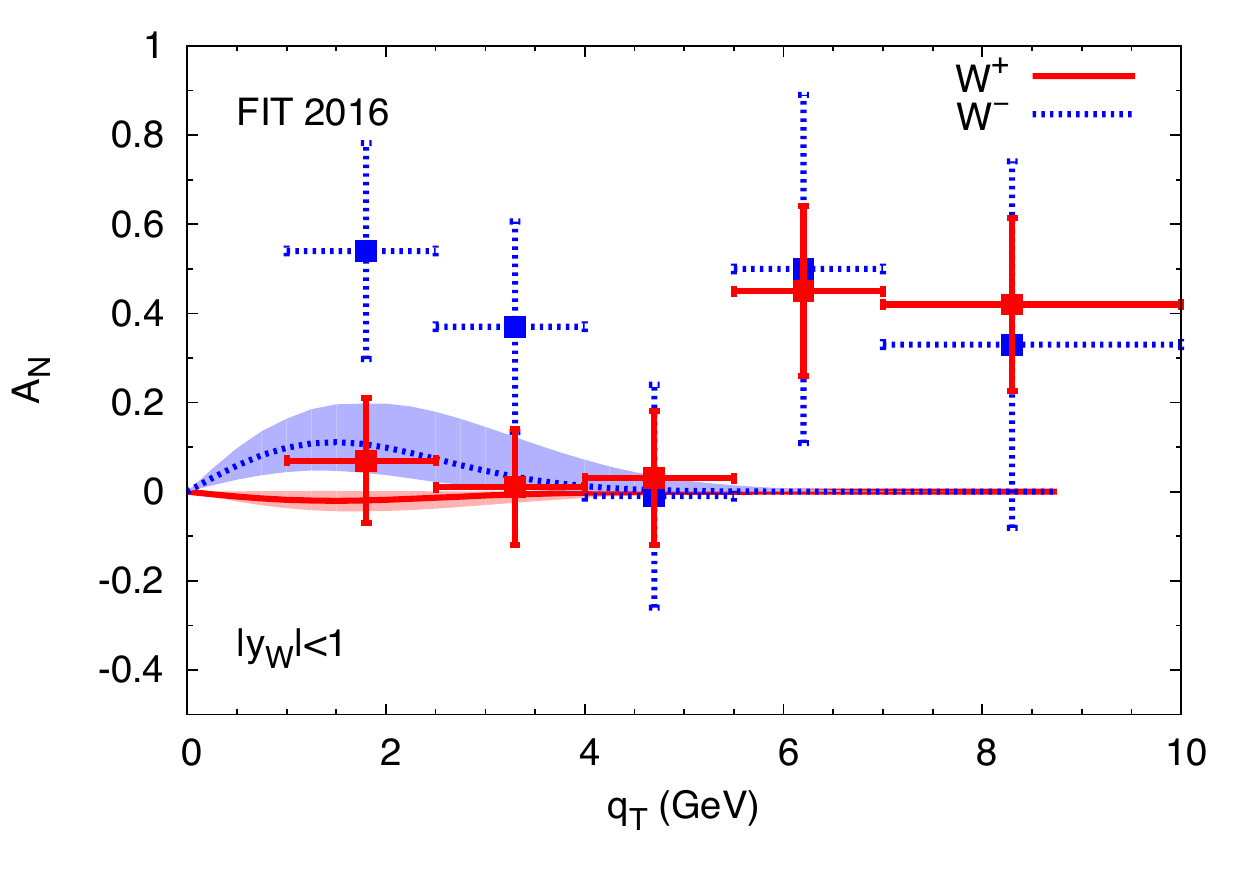}
\caption{Our estimates of the Sivers asymmetry $A_N$ for $W^+$ and $W^-$
production, assuming a sign change of the SIDIS Sivers functions, compared
with the experimental data as function of $q_T$. $y_{_W}$ is integrated in the
region $[-1, 1]$.}
\label{fig:sivers_W_y}
\end{figure}

\begin{figure}[htbp]
\centering
\includegraphics[width=9cm]{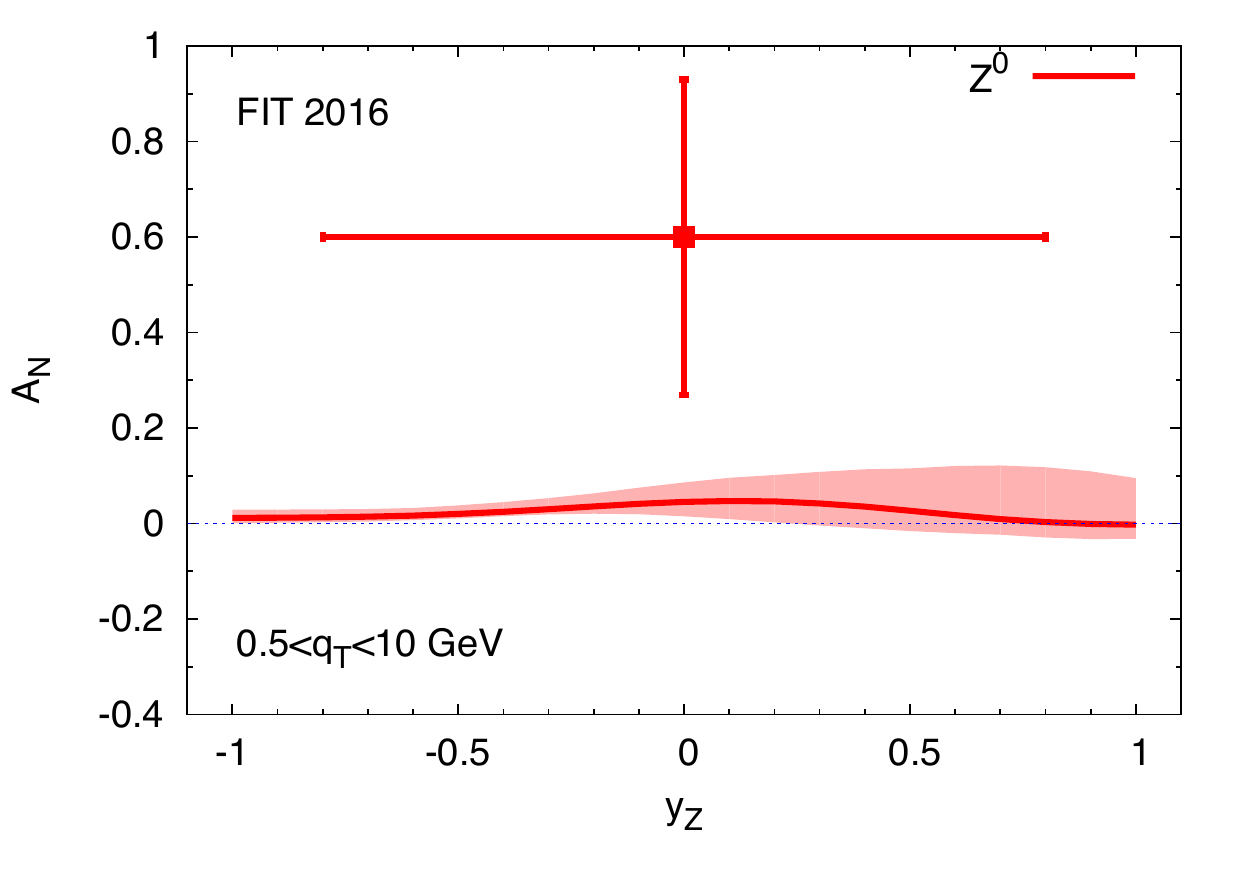}
\caption{Our estimate of the Sivers asymmetry $A_N$ for $Z^0$
production, assuming a sign change of the SIDIS Sivers functions, compared
with the experimental data as function of $y_{_Z}$. $q_T$ is integrated in the
region $[0.5, 10]$ GeV.}
\label{fig:sivers_Z}
\end{figure}

Actually, in order to compare with data~\cite{Adamczyk:2015gyk}, we integrate
both the numerator and denominator of $A_N^W$, Eq.~(\ref{ann}), either over
$q_T$ in the region $[0.5, 10]$ GeV, or over $y_{_W}$ from $-1$ to 1. The
results, {\it reversing the sign of the SIDIS extracted Sivers functions} as
in Eq.~(\ref{eq:sign-change}), are shown and compared with data respectively
in Fig.~\ref{fig:sivers_W} and in Fig.~\ref{fig:sivers_W_y}. For completeness,
despite the much limited amount and quality of data, we also show our estimate
of $A_N$, integrated over $q_T$, for $Z^0$ production, in Fig.~\ref{fig:sivers_Z}.
The results {\it without the sign change} can be easily deduced by reversing
the sign of the asymmetry in Figs.~\ref{fig:sivers_W}-\ref{fig:sivers_Z}.


Before trying, in the next Section, a quantitative evaluation of the
significance of the data regarding the issue of the sign change of the Sivers
function going from SIDIS to D-Y processes, a few comments are in order.
\begin{itemize}
\item
In general, the agreement between our estimates and the few data is rather
poor, both with and without sign change. In particular, this is evident from
the $q_T$ dependence of $A_N$, Fig.~\ref{fig:sivers_W_y}, and the $y_{_Z}$
dependence of $A_N$ for $Z^0$, Fig.~\ref{fig:sivers_Z}. In the latter
case there is only one single data point, with a big error, indicating
a large positive asymmetry.
\item
The data on the $y_{_W}$ dependence are given by collecting all $W$'s produced
with $q_T$ up to 10 GeV. The simple model of D-Y TMD factorisation without
evolution that we use in this analysis is expected to hold for lower values
of $q_T$; integrating the theoretical results up to such values, in order to
compare with the available data, is a somewhat ambiguous procedure.
Implementation of the TMD evolution would not help to make the agreement
with the data better in this case, as TMD evolution predicts a suppression
of the asymmetries for higher values of $Q^2$ with respect to the initial
lower scale~\cite{Echevarria:2014xaa}. This suppression might become moderate
depending on the shape of the non-perturbative input of
TMD evolution \cite{Aidala:2014hva,Collins:2014jpa,Kang:2015msa}.
\item
Considering the $q_T$ integrated data, from a first look at Fig.~\ref{fig:sivers_W}
it appears that indeed $W^-$ data are compatible with the sign change, while
$W^+$ data may be compatible with either sign of the Sivers functions.
\item
The shape of the TMDs and the values of the parameters here adopted allow
a good description of the SIDIS data; however, they are still rather flexible,
and our numerical estimates for the D-Y asymmetry might depend on the choice,
for example, of the values of the Gaussian width, Eq.~(\ref{hermes-par}).
A full study of combined unpolarised  SIDIS, D-Y and (future) $e^+e^-$
data is mandatory.
\end{itemize}

\section{\label{Imp} Impact of the asymmetries on the extraction of Sivers functions}

\begin{figure}[tbp]
\centering
\includegraphics[width=6.5cm]{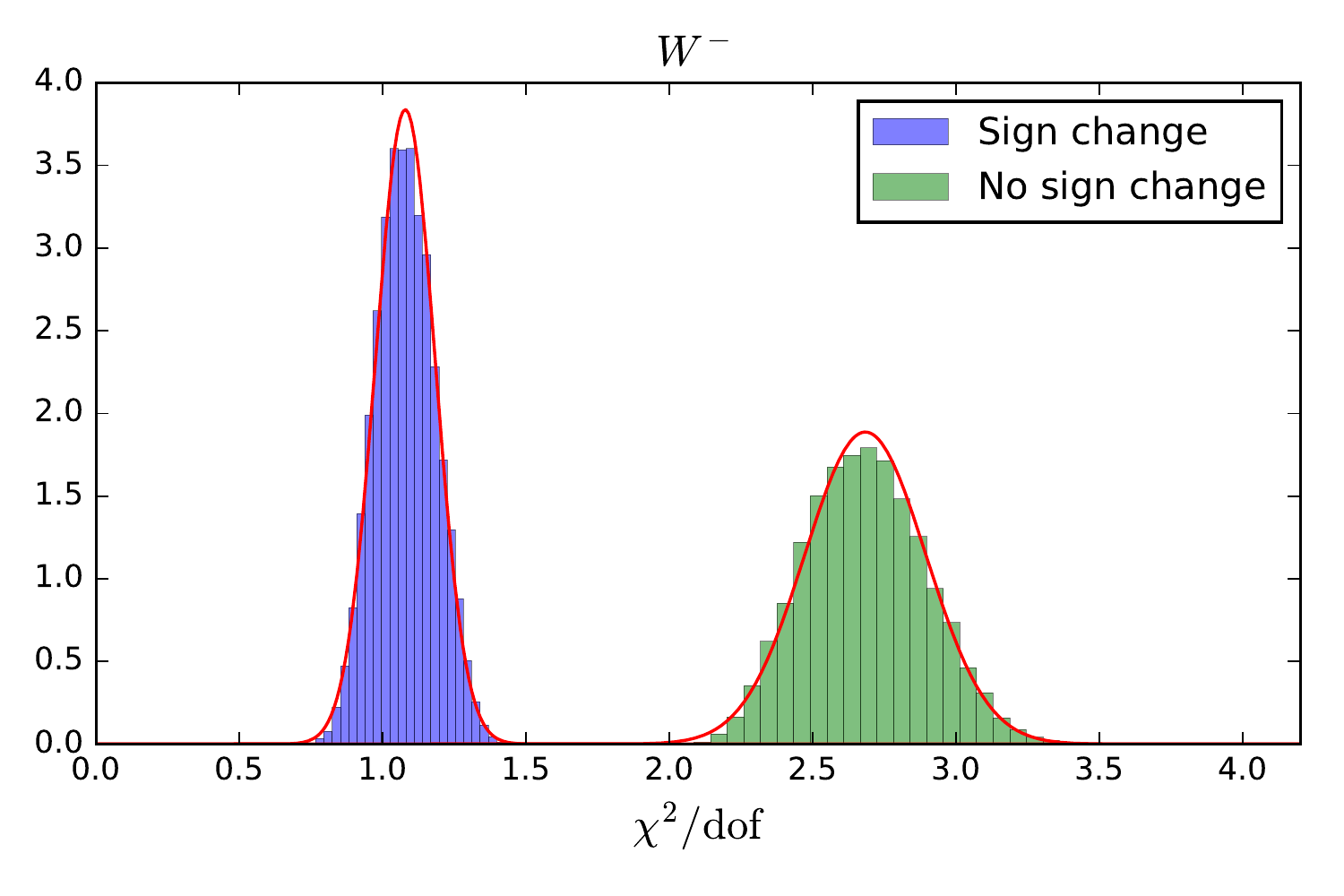} (a)
\includegraphics[width=6.5cm]{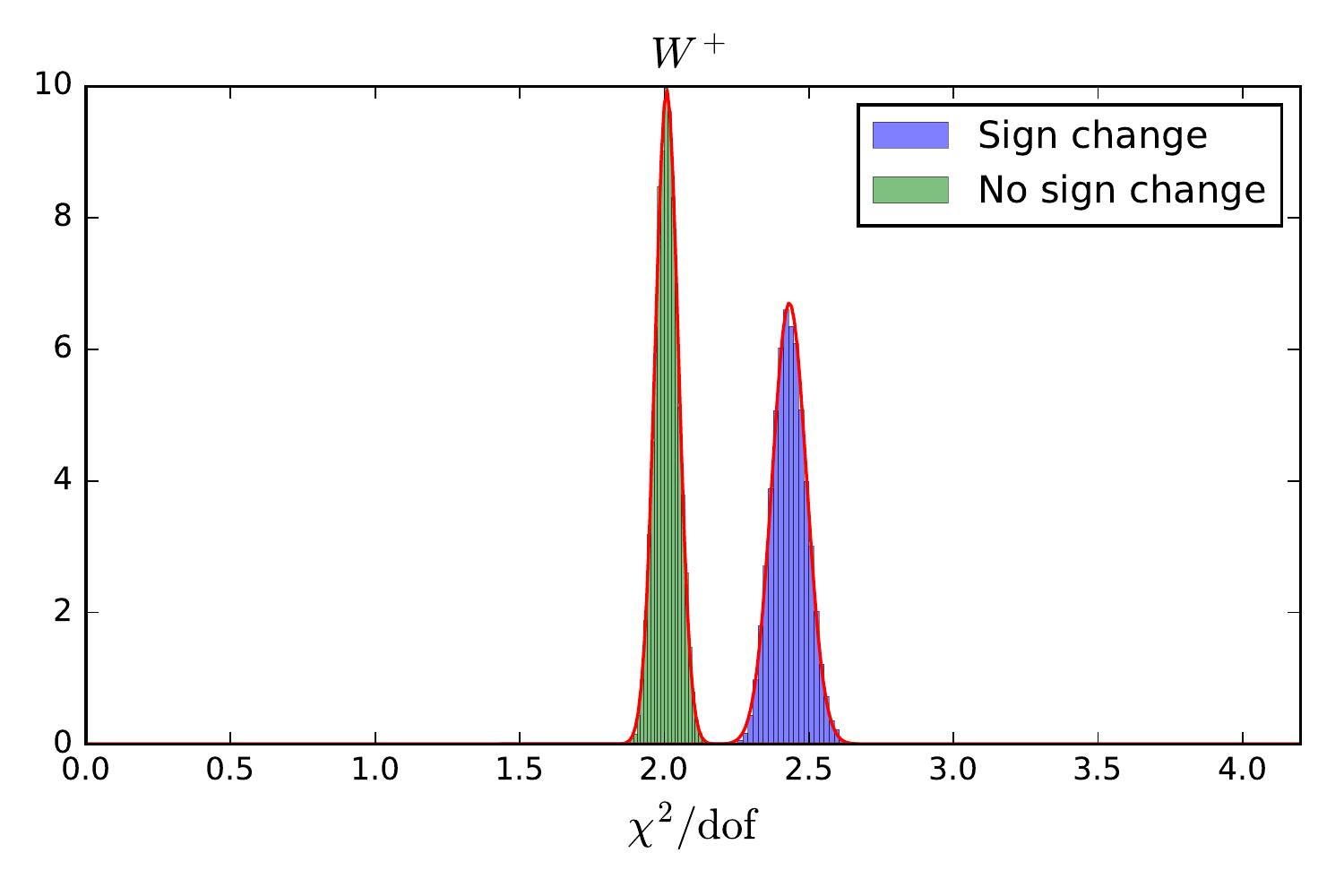} (b)
\includegraphics[width=9cm]{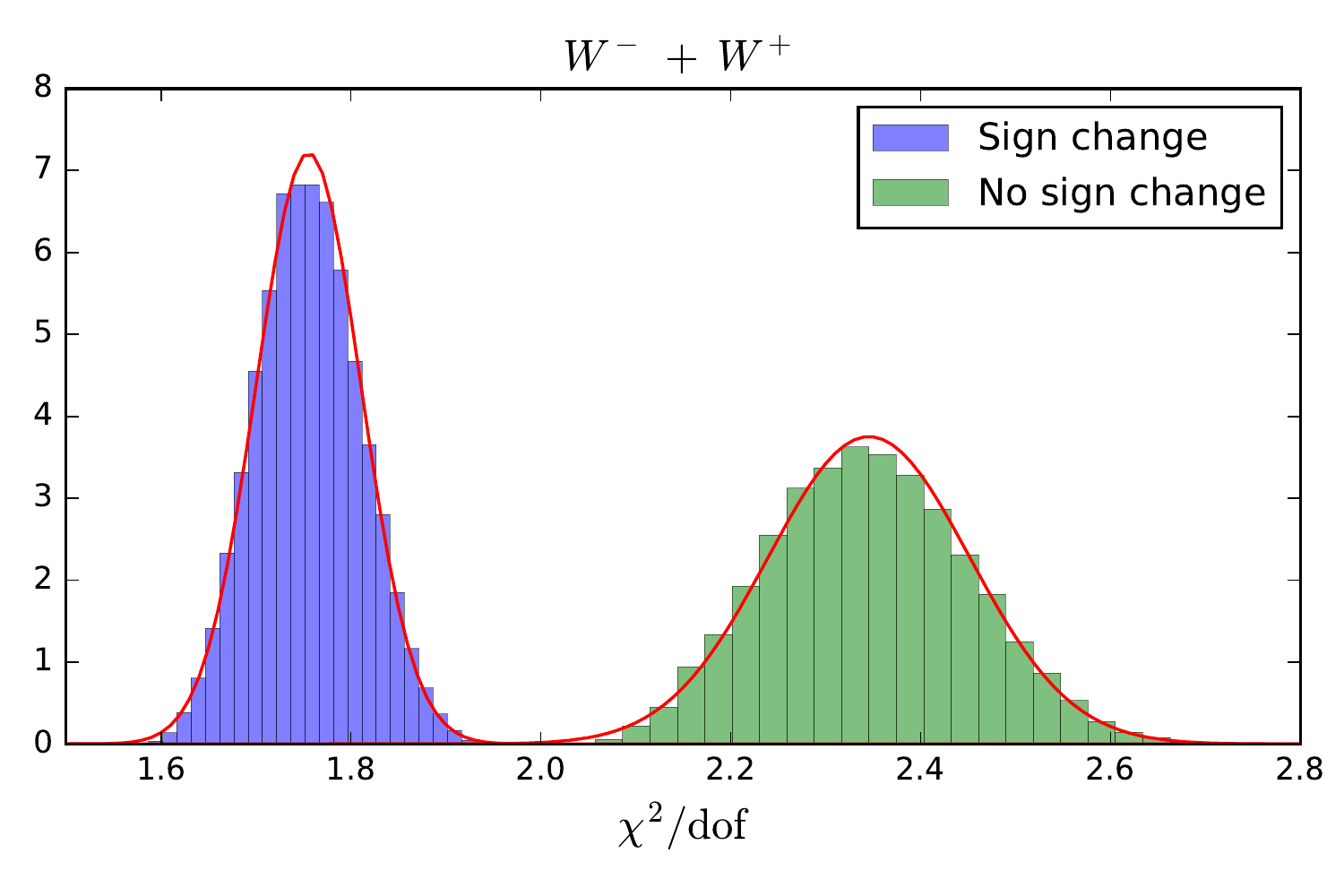} (c)
\caption{{\bf a,b:} Probability density functions of the
$\chi^2/{\rm dof}$ separately for our predictions of $W^-$ (left)
and $W^+$ (right) asymmetries, obtained from all parameter sets used
to calculate the error band. The green histograms correspond to no sign
change of the Sivers function, while the blue histograms correspond to
the sign change. Fitted normal distributions are shown as solid lines.
{\bf c:} Probability density functions of the $\chi^2/{\rm dof}$,
as in the upper plots, but globally for our predictions of $W^-$ + $W^+$
asymmetries.}
\label{fig:sivers_sign_change}
\end{figure}

In this Section we take at face value the RHIC data on $A_N$ for $W^\pm$
production and, in order to quantify their significance, we calculate the
deviation between the data and our estimates, separately for $W^+$ and $W^-$:
\be
\chi^2(\bm{\alpha}) =  \sum_{n=1}^{\rm dof} \left(
\frac{[{\rm theory}]_n(\bm{\alpha})
-[{\rm exp}]_n}{[\Delta{\rm exp}]_n}\right)^2 \, \label{eq:chi2}
\ee
where $[{\rm theory}]_n(\bm{\alpha})$ corresponds to the calculation of
the $W$ asymmetry using the phenomenological extraction of the Sivers function
performed in this paper, with model parameters $\bm{\alpha}$, with and
without the sign change of Eq.~(\ref{eq:sign-change}); $[{\rm exp}]_n$ are
the data for $W^+$ or $W^-$ asymmetries and $[\Delta{\rm exp}]_n$ are
the corresponding experimental errors. As we explained in Section~\ref{Extr},
in order to estimate the error on the extraction of the Sivers functions, we
generate 2$\cdot 10^4$ sets of parameters $\bm{\alpha}$ according to
Eq.~(\ref{condition}).
Thus, we calculate 2$\cdot 10^4$ values of $\chi^2$ using Eq.~(\ref{eq:chi2})
for $W^+$ and $W^-$. The histogram of all these values of $\chi^2/{\rm dof}$
are shown in Fig.~\ref{fig:sivers_sign_change}, where ${\rm dof} = 8$ is the
number of experimental points in each set for $W^\pm$. The green histogram
corresponds to $\chi^2$ with no sign change of the Sivers function, while
the blue histogram corresponds to $\chi^2$ with the sign change of the Sivers
functions.

One can see from the upper left panel of Fig.~\ref{fig:sivers_sign_change}
that $W^-$ data favour the sign change: in this case the values of
$\chi^2/{\rm dof}$ are around 1.1, while without the sign change they are
around 2.7. The $W^+$ data on the other hand are slightly better with no
sign change, as can be seen from the upper right panel of
Fig.~\ref{fig:sivers_sign_change}. For either scenarios the $\chi^2$ per
number of data are rather large: these large values are due to the single
point at $y_{_W}=0$ (see Fig.~\ref{fig:sivers_W}, left panel) and the two
points at large $q_T > 5$ GeV (see Fig.~\ref{fig:sivers_W_y}).

If we combine both $W^+$ and $W^-$ data, then the two data sets globally favour
a sign change of the Sivers functions according to Eq.~(\ref{eq:sign-change}).
The histogram of the combined data sets is presented in the lower panel of
Fig.~\ref{fig:sivers_sign_change}. If one assumes no sign change, then
$\langle \chi^2/{\rm dof} \rangle = 2.35$  and $\sigma( \chi^2/{\rm dof} )
= 0.1$, where dof = 16, while the sign change yields a
lower value, $\langle \chi^2/{\rm dof} \rangle = 1.75$ and
$\sigma( \chi^2/{\rm dof}) = 0.05$. Notice that both
scenarios have some disagreement with our estimates: indeed the values of
$\chi^2/{\rm dof}$ are well above one. Using our results from
Fig.~\ref{fig:sivers_sign_change} we can at most conclude that $W^\pm$
data hint at an indication of the sign change according to
Eq.~(\ref{eq:sign-change}).

Another interesting question that we would like to investigate in this paper
is whether the data on the $W^\pm$ asymmetries have any significant impact
on the parameters of our model. Notice that we do not include $W^\pm$ data
in our fit of the Sivers functions. Bayes theorem allows to incorporate
information from new data by applying a re-weighting of the probability
densities for the model parameters. The details of application of the
re-weighting are explained in Ref.~\cite{Sato:2013ika}. The probability
density function for model parameters $\bm{\alpha}$, $\mathcal{P}(\bm{\alpha})$,
is going to be modified in presence of new data, and the Bayes theorem states
that
\begin{align}
\mathcal{P}(\bm{\alpha}|D)
=\frac{\mathcal{P}(D|\bm{\alpha})}{\mathcal{P}(D)} \mathcal{P}(\bm{\alpha}),
\label{eq:bayes}
\end{align}
where $\mathcal{P}(\bm{\alpha}|D)$ is the so-called \emph{posterior} density,
that is the updated probability density function from the \emph{prior} density
$\mathcal{P}(\bm{\alpha})$. The quantity $\mathcal{P}(D|\bm{\alpha})$, called
the \emph{likelihood} function, represents the conditional probability for a
data set $D$ given the parameters $\bm{\alpha}$ of the model.
The quantity $\mathcal{P}(D) $ ensures the normalisation of the
posterior density to unity.

For a particular observable $\mathcal{O}$ one can write the expectation value
with the new data as,
\begin{align}
\text{E}[\mathcal{O}]
=	\int \! d^n\!\alpha \, \mathcal{P}(\bm{\alpha}|D) \,
	\mathcal{O}(\bm{\alpha}) .
\label{eq:E}
\end{align}
In order to estimate the integral of Eq.~(\ref{eq:E}) we will use a Monte Carlo
approximation of the integral, such that the integral over
continuous values of $\bm{\alpha}$ will be substituted with a sum over discrete
values of $\bm{\alpha}_k$. These $\bm{\alpha}_k$ are the generated
2$\cdot 10^4$ sets of parameters according to Eq.~(\ref{condition}). We obtain
\begin{align}
\text{E}[\mathcal{O}]
\simeq	\frac{1}{N} \sum_k w_k \, \mathcal{O}(\bm{\alpha}_k),
\label{eq:E1}
\end{align}
where the quantities $w_k$ are called \emph{weights} and are proportional to
$\mathcal{P}(D|\bm{\alpha}_k)$.
Their normalisation is fixed by demanding $\text{E}[1]=1$, that is,
$\sum_k w_k=N$.

Similarly, the variance  of an observable $\mathcal{O}$ is given by
\begin{align}
\text{Var}[\mathcal{O}]
&=\frac{1}{N} \sum_k w_k \left( \mathcal{O}(\bm{\alpha}_k)-\text{E}
[\mathcal{O}] \right)^2 .
\label{eq:Var}
\end{align}
%

%
\begin{figure}[htbp]
\centering
\includegraphics[width=6cm]{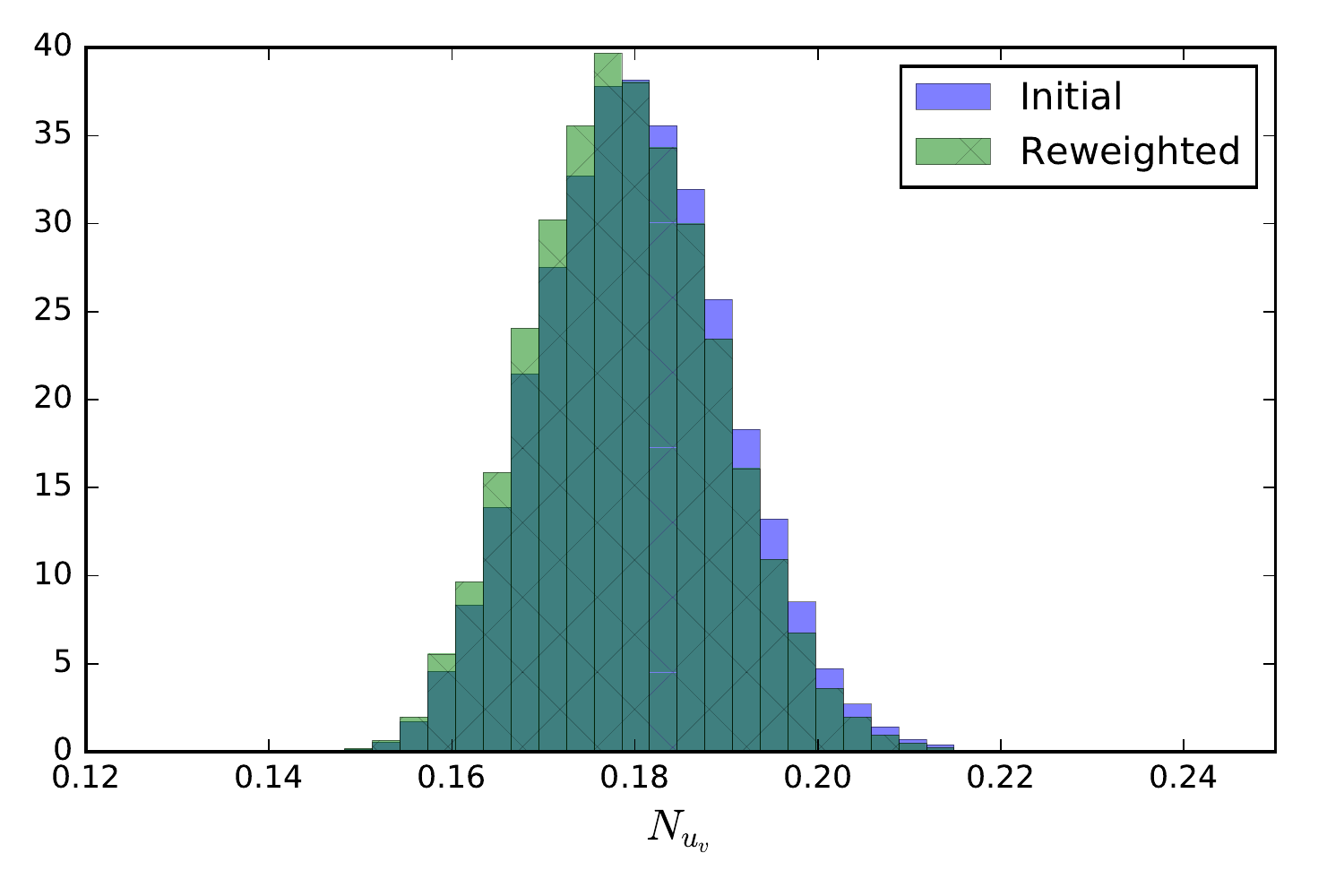}
\includegraphics[width=6cm]{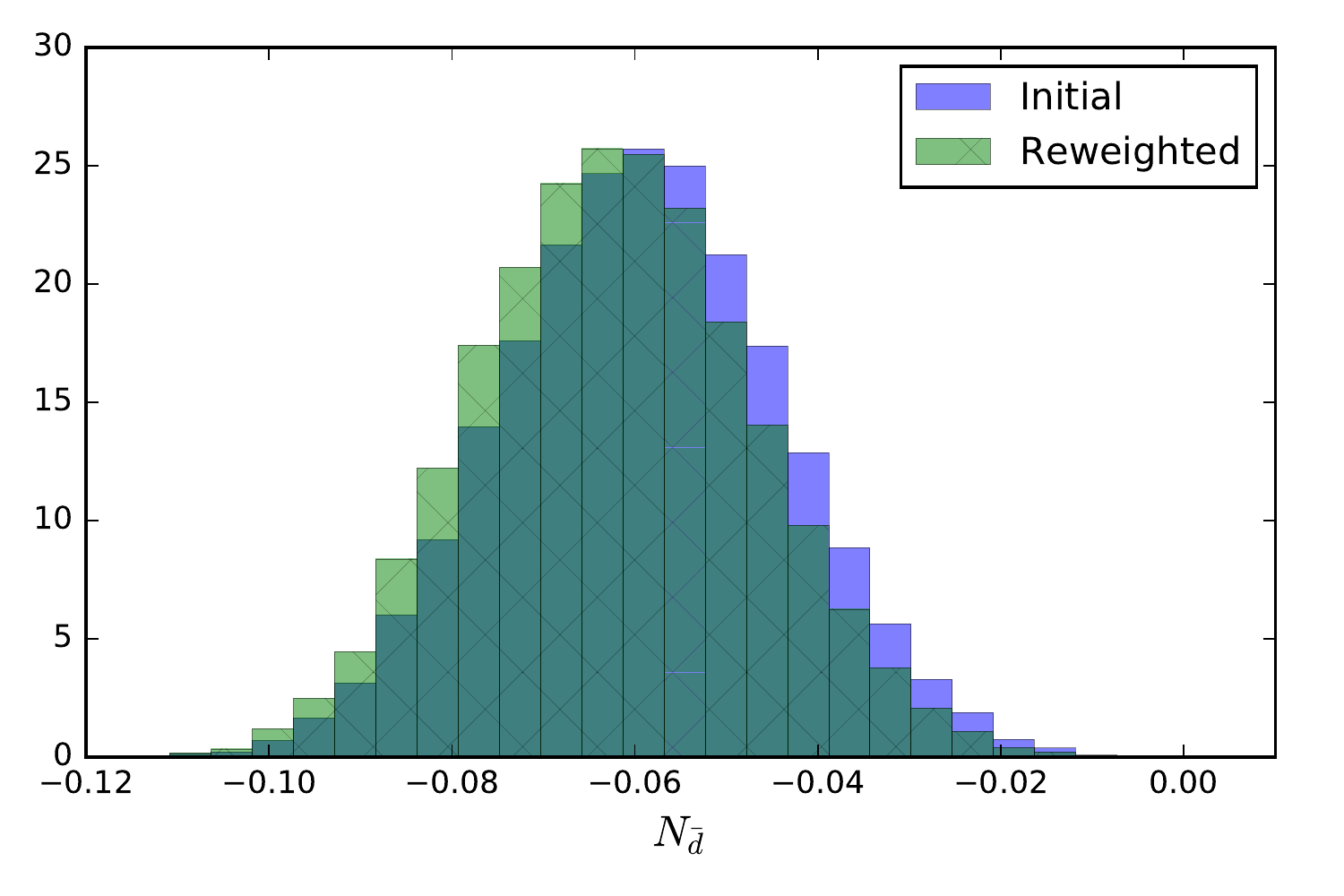}
\includegraphics[width=6cm]{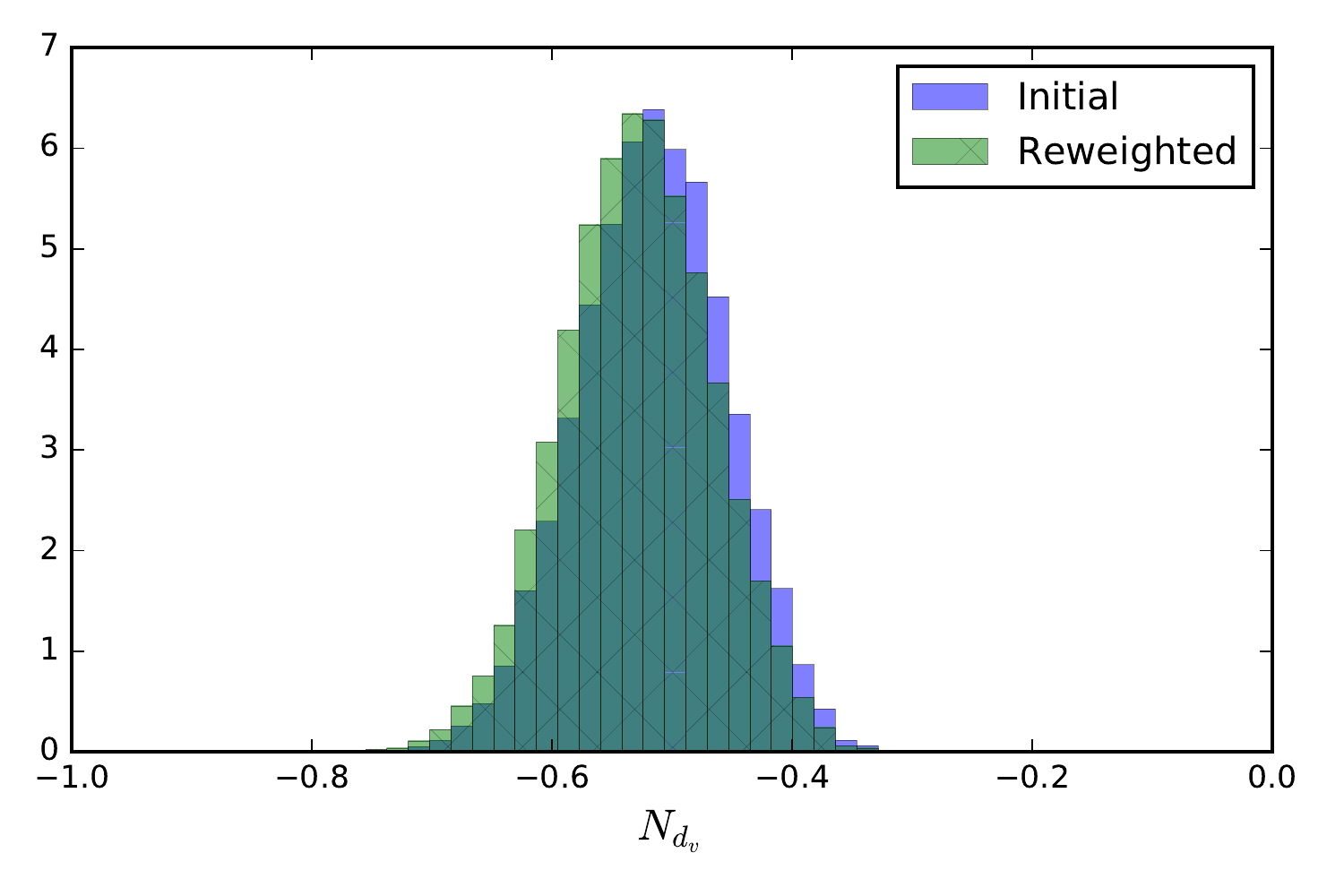}
\includegraphics[width=6cm]{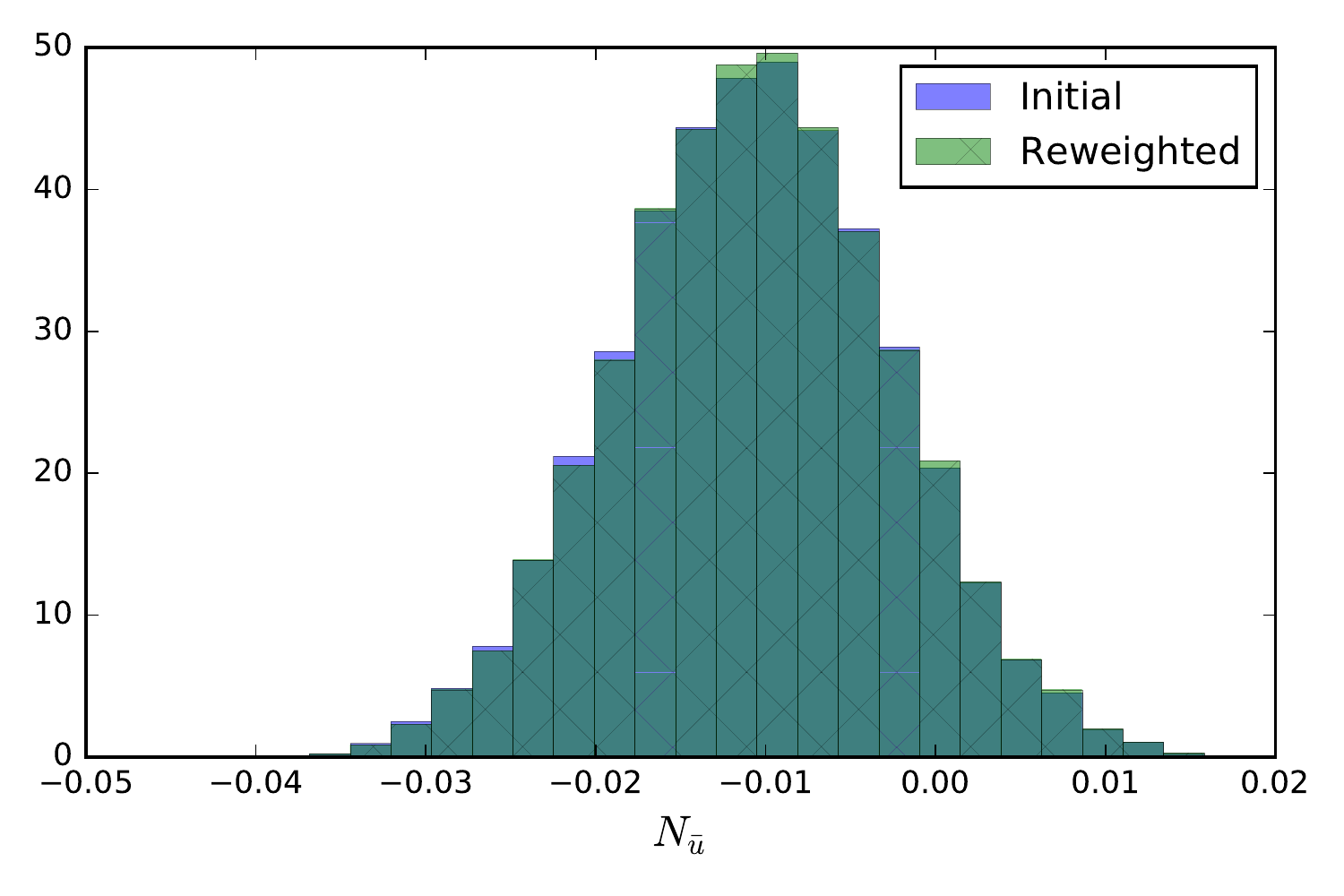}
\caption{The re-weighting procedure applied to $N_{u_v}$, $N_{\bar d}$, $N_{d_v}$, and $N_{\bar u}$ parameters.
The green hatched histogram corresponds to re-weighting, while the blue
histogram corresponds to the prior distribution.}
\label{fig:sivers_sets}
\end{figure}

The re-weighting procedure depends on the form assumed for the likelihood
function. We use $\chi^2$ minimisation in the fits, then our weights
have the following form:
\begin{align}
	w_k \propto \exp\left(-\frac{1}{2}\chi^2(\bm{\alpha}_k)\right),
\label{eq:bayes1}
\end{align}
where all values of $\chi^2(\bm{\alpha}_k)$ can be readily obtained from
our results and Eq.~(\ref{eq:chi2}). We have checked that the form of the
\emph{prior} density $\mathcal{P}(\bm{\alpha})$ for our parameters turns
out to be very well approximated by the normal distribution.

The re-weighting can be readily applied to all parameters of the model, as well
as to observables: in Fig.~\ref{fig:sivers_sets} we show, as an example, the
application of the re-weighting procedure to the parameters that describe the
normalisation of the $u$-valence, $d$-valence, anti-$d$, and anti-$u$ quark
Sivers functions. One can see that the re-weighted densities (green hatched
histograms) are only slightly shifted from the prior distributions (blue
histograms), and the new mean value is within $1\sigma$ of the current values,
given in  Table~\ref{parameters}. The same observation is true for all other
parameters.

\section{\label{Com} Comments and conclusions}
We have analysed the recent data on the single spin asymmetry $A_N^W$ measured
by the STAR Collaboration at RHIC~\cite{Adamczyk:2015gyk}; it is the first
ever spin asymmetry measured in Drell-Yan processes and it might originate
from the fundamental Sivers distribution of polarised quarks in an
unpolarised proton. Then, it could help in testing the validity of the widely
expected sign change of the Sivers function when extracted in SIDIS and D-Y
processes, Eq.~(\ref{eq:sign-change}).

In order to perform an unbiased analysis we have re-derived, by best fitting
the latest SIDIS data~\cite{Airapetian:2009ae,Alekseev:2008aa,Adolph:2014zba,Qian:2011py},
the Sivers functions, including the anti-quark ones
which might play a role in the D-Y production of $W$s and $Z^0$s.
Our results are shown in Tables~\ref{parameters} and~\ref{results_sidis} and in
Fig.~\ref{fig:functions}.

Using the newly extracted Sivers SIDIS functions we have computed the D-Y SSA
$A_N$ for $W^\pm$ and $Z^0$ production, both with and without a sign change
of the Sivers functions. Then, we have compared our results with the STAR data,
trying to assess their significance with respect to the sign change issue.
Our quantitative results, according to Eq.~(\ref{eq:chi2}), can be seen in
Fig.~\ref{fig:sivers_sign_change}.

As commented throughout the paper, our simple model of D-Y TMD factorisation
without evolution, Eqs.~(\ref{asy})-(\ref{siverskt}), is, in general, in poor
agreement with the data. A more refined analysis, using the TMD evolution,
would probably worsen the agreement~\cite{Echevarria:2014xaa}. One should
add that the data, although important and pioneering, are still scarce, with
large errors, and gathered in different kinematical regions.

With all the necessary caution, from our analysis of the data, one can at most
conclude that, only from $W^-$ production, there is an indication in favour of
the sign change of the Sivers function, which, however, is still far from
being considered
as proven. Soon expected data from COMPASS polarised D-Y processes,
$\pi^- p^\uparrow \to \ell^+ \ell^- X$, and higher statistics data from STAR Collaboration on $W$ and $Z$ production should add important information.

\acknowledgments
\noindent
This work was partially supported by the U.S.~Department of Energy, Office of Science, Office of Nuclear Physics, within the framework of the TMD Topical Collaboration, and under Contract No.~DE-AC05-06OR23177 (A.P.)  and by the National Science Foundation under Contract No. PHY-1623454 (A.P.).
M.A. and M.B.~acknowledge support from the ``Progetto di Ricerca Ateneo/CSP"
(codice TO-Call3-2012-0103).

%
%

\providecommand{\href}[2]{#2}\begingroup\raggedright\endgroup

\end{document}